\documentclass[useAMS,usenatbib]{mnras}

\usepackage{kantlipsum,widetext}
\usepackage{times,epsfig}
\usepackage{graphics,graphicx}
\usepackage{amsmath,amsfonts}
\usepackage{epsfig}
\usepackage{url}
\usepackage{morefloats}
\usepackage{tabularx}
\usepackage{color}
\usepackage{ae,aecompl}

\usepackage{calligra}
\usepackage{inslrmaj}

\usepackage[caption=false]{subfig}

\usepackage{mathrsfs}
\usepackage{amssymb}
\usepackage{bm}

\usepackage{natbib}
\def\thetaB{\mbox{\boldmath$\theta$}}

\def\ba#1\ea{\begin{align}#1\end{align}}

\newcommand{\<}{\langle}
\renewcommand{\>}{\rangle}

\renewcommand{\d}{\delta}
\newcommand{\D}{\Delta}
\renewcommand{\l}{\ell}

\newcommand{\zt}{\tilde{z}}

\newcommand{\chit}{\tilde{\chi}}
\newcommand{\vs}{\nonumber\\}
\renewcommand{\d}{\delta}

\definecolor{darkred}{rgb}{0.8,0,0}

\def\lsim{\,\lower2truept\hbox{${<\atop\hbox{\raise4truept\hbox{$\sim$}}}$}\,}
\def\gsim{\,\lower2truept\hbox{${> \atop\hbox{\raise4truept\hbox{$\sim$}}}$}\,}
\def\simlt{\mathrel{\rlap{\lower 3pt\hbox{$\sim$}}
        \raise 2.0pt\hbox{$<$}}}
\def\simgt{\mathrel{\rlap{\lower 3pt\hbox{$\sim$}}
        \raise 2.0pt\hbox{$>$}}}

\def\be{\begin{equation}}
\def\ee{\end{equation}}
\def\ba{\begin{eqnarray}}
\def\ea{\end{eqnarray}}

\def\l{\left}
\def\r{\right}

\def\be{\begin{equation}}
\def\ee{\end{equation}}

\def\hksqrt{\mathpalette\DHLhksqrt}
\def\DHLhksqrt#1#2{\setbox0=\hbox{$#1\sqrt{#2\,}$}\dimen0=\ht0
\advance\dimen0-0.2\ht0
\setbox2=\hbox{\vrule height\ht0 depth -\dimen0}%
{\box0\lower0.4pt\box2}}

\title[GW astronomy with radio surveys]{Gravitational wave astronomy with radio galaxy surveys}
\author[Alvise Raccanelli]
{\parbox[t]{\textwidth}
{Alvise Raccanelli}
\vspace*{8pt}\ \\
Department of Physics {\fontfamily{ppl}\selectfont \&} Astronomy, Johns Hopkins University, Baltimore, MD 21218 USA}
\date{}
\begin{document}
\maketitle

\begin{abstract}
In the next decade, new astrophysical instruments will deliver the first large-scale maps of gravitational waves and radio sources. Therefore, it is timely to investigate the possibility to combine them to provide new and complementary ways to study the Universe.
Using simulated catalogues appropriate to the planned surveys, it is possible to predict measurements of the cross-correlation between radio sources and GW maps and the effects of a stochastic gravitational wave background on galaxy maps.
Effects of GWs on the large scale structure of the Universe can be used to investigate the nature of the progenitors of merging BHs, the validity of Einstein's General Relativity, models for dark energy, and detect a stochastic background of GW.
The results obtained show that the galaxy-GW cross-correlation can provide useful information in the near future, while the detection of tensor perturbation effects on the LSS will require instruments with capabilities beyond the currently planned next generation of radio arrays.
Nevertheless, any information from the combination of galaxy surveys with GW maps will help provide additional information for the newly born gravitational wave astronomy.
\end{abstract}



\section{Introduction}
The detection by the LIGO instrument of gravitational waves (GW150914,~\citealp{GW150914} and GW151226,~\citealp{GW151226}) from the merger of binary black holes opened up a new window to study our Universe. In the first few months following the first detection, gravitational waves have been used to test General Relativity in a new way~\citep{LIGO:GR}, the speed of gravitational waves~\citep{Collett:2016} and alternative cosmological models such as the one where the dark matter is made of primordial black holes (e.g.~\citealt{Bird:2016}).

Currently and for the foreseeable future, the main way to detect gravitational waves (GWs) is by the use of laser interferometers, on Earth and in space. Several alternatives have been proposed, and they involve detecting the effect of GWs on other observables, such as Pulsar timing arrays~\citep{PTA}, the effect of gravitational waves from inflation on the Cosmic Microwave Background~\citep{KK}, and the effect of GWs on the Large-Scale Structure (LSS) of the Universe~\citep{Guzzetti:2016}.

The presence of tensor modes during the early epochs of the Universe modifies the power-spectrum of primordial scalar perturbations~\citep{Jeong:2012CF}, while at late times the presence of a GW background leads to several effects, including projection effects due to the perturbation of space-time by GWs on the galaxy distribution~\citep{Jeong:2012, Schmidt:2012}, the CMB~\citep{Dodelson:2003, Cooray:2005, Book:2011a} and the $21$-cm background~\citep{Book:2012, Pen:2003}.

At the same time, radio surveys for cosmology are entering a new phase of exponential expansion on both quantity and quality of data available~\citep{sparcs}, with the construction of several instruments, including the Australian Square Kilometre Array Pathfinder (ASKAP,~\citealt{Johnston:2008}) and the design definition of the Square Kilometre Array (SKA\footnote{https://www.skatelescope.org}).
Radio galaxy surveys with such instruments will be able to detect galaxies over a large redshift range, a wide area of the sky, and down to a very low flux limit.

Radio surveys such as NVSS have been used in the past to perform cosmological analyses (see e.g.~\citealt{Nolta:2004, Raccanelli:2008, Xia:2010, Bertacca:ISW}); future surveys will have a wider redshift range and orders of magnitude more objects observed, so it is expected they will improve the precision of cosmological measurements~\citep{Raccanelli:radio}.
All this, combined with the fact that effects of GWs on LSS are largest at very large scales, makes it very timely to start an investigation of the combination of GW with radio galaxy maps.

This paper investigates the possibility to use future radio galaxy surveys to contribute to gravitational wave astronomy. By measuring the position and correlation of galaxies, or cross-correlating their number counts with GW maps, it will be possible to detect direct or indirect effects of GWs coming from the merger of massive compact objects or the early stages of the Universe.

Gravitational wave astronomy is still in its infancy but it is predicted to grow quickly, and the coincidental exponential increase in radio survey capabilities makes it very interesting to analyze how to best combine the two fields.
Therefore, it is timely to try to understand if the combination of observations of radio sources and GWs can give useful additional information about cosmological models and parameters currently investigated.
%

We will present forecasts of the constraints on cosmological models and parameters that will be possible to obtain both by cross-correlating future GW maps with galaxy catalogs from a variety of planned radio surveys, and by analyzing the effect of GWs on position, distribution and correlation of such galaxy catalogs.
Recently, ideas about cross-correlation of LSS with GW maps have been explored in e.g.~\cite{Camera:2013, Oguri:2016, Namikawa:2016, Raccanelli:2016PBH}.

The structure of the paper is as follows. In Section~\ref{sec:radiosurveys} we introduce the radio galaxy surveys we consider.
In Section~\ref{sec:ClgGW} we present the studies that will be enabled by cross-correlating GW with galaxy maps, in particular angular correlations to determine the progenitor of BBH mergers in Section~\ref{sec:ccf} and constraints on cosmic acceleration models by using lensing effects on radial correlations in Section~\ref{sec:cosmag}.
In Section~\ref{sec:SGWB} we investigate the effects of GWs on the LSS; we predict measurements that will be possible to obtain by using cosmometry in Section~\ref{sec:cosmometry} and the cosmic rulers methodology in Section~\ref{sec:rulers}.
We then summarize our findings and conclude in Section~\ref{sec:conclusions}.

\section{Forthcoming radio galaxy surveys}
\label{sec:radiosurveys}
In this paper we focus on forthcoming radio galaxy surveys; we consider the Square Kilometre Array (SKA,~\citealt{SKA:Maartens}) and its pathfinders~\citep{sparcs}, in particular the cosmology survey Evolutionary Map of the Universe (EMU,~\citealt{emu}) with the ASKAP instrument. We will consider both radio continuum and H{\sc i} spectroscopic cases.

We model radio surveys by using the prescription of~\cite{Wilman:2008}; catalogues are generated from the S-cubed simulation\footnote{http://s-cubed.physics.ox.ac.uk}, using the SEX and SAX for continuum and H{\sc i} surveys, respectively. We then apply a cut to the simulated data to reflect the assumed flux limit for different cases.
More details on the underlying modeling and planned surveys can be found in~\cite{Wilman:2008, SKA:continuum, SKA:HI}.
Finally, for all surveys we assume $f_{\rm sky} =0.75$.
We expect that for some measurements, in particular the detection of GW effects using only radio surveys, the area, redshift range and number density required will often be surpassing the specifications of planned instruments. Therefore, we include in our predictions instruments with futuristic capabilities, which will be used to understand if those measurements will be possible even in principle.

Although radio continuum surveys do not have in principle redshift information, there are a variety of techniques that could enable the possibility to divide the galaxy catalog into redshift bins.
A detailed comparison of methodologies is beyond the scope of this paper, where we assume the clustering-based redshift (CBR) information proposed in~\cite{Menard:2013}.
We will assume the possibility to divide the catalog into bins using the CBR technique following~\cite{Kovetz:2016}.

\subsection{EMU}
At its completion, ASKAP will consist of 36 12-meter antennas spread over a region 6 km in diameter.
EMU is an all-sky radio continuum survey that will cover the whole southern sky, extending as far north as +30 deg., with a sensitivity of 10$\mu{\rm Jy}$/beam rms, over a frequency range of 1130-1430 MHz.
It will be a particularly fast survey, thanks to the phased-array feed (PAF) at the focus of each antenna, which gives ASKAP a 30 sq. deg. of instantaneous field of view.
It is expected to detect $\sim70$ million galaxies, which will make it the largest radio galaxy survey so far.

\subsection{SKA}
The Square Kilometre Array (SKA) is an international multi-purpose next-generation radio interferometer, that will be built in phases (SKA1 and SKA2) in the Southern Hemisphere in South Africa and Australia; it will cover a frequency range of 350 MHz - 15 GHz, have a central core of $\sim200$ km diameter, with 3 spiral arms spreading over 3000 km and a total collecting area of about 1 km$^2$.
SKA is a facility that is planned to last for around 50 years, and it will continuously scan the sky, producing a vast amount of data.
Among many types of observations delivered by such instruments, we focus here on surveys that will detect individual galaxies, neglecting e.g. intensity mapping surveys.
On longer time-scales, one can envisage significant improvements with respect to the instruments planned at the moment.

\subsubsection{Radio continuum}
Being unaffected by dust, radio continuum emissions can be used to detect star forming galaxies and AGN up to very high redshift.
Radio continuum surveys with the SKA will observe 30,000 deg.$^2$ out to extremely high redshift, detecting $\sim10^8$ galaxies
in SKA1 and $\sim10^9$ in SKA2.

In the top panel of Figure~\ref{fig:Nz_radio} we show the predicted redshift distributions for continuum surveys.
We show three different flux limits $S_{\rm lim}$: (i) 50$\mu$Jy, corresponding to a 5-$\sigma$ detection for sources with the planned full EMU survey; (ii) an {\it SKA-like} 1$\mu$Jy limit, and (iii) a futuristic distribution including all sources down to 1$n$Jy. In the legend we also include the total number of objects per square degree, that corresponds to $\approx$ 8500, 135000, 550000, respectively.

\subsubsection{H{\sc i}}
While current H{\sc i} galaxy surveys are not competitive with wide optical ones, technological advancements that will be implemented in the SKA will allow for dramatically faster survey speeds, making it possible to map the galaxy distribution out to high redshifts over large areas, therefore enabling measurements of ultra-large scales.
Once the full SKA2 will be in place, it will provide what has been called the ``billion galaxy survey'', that will be the biggest spectroscopic galaxy survey to date, detecting $\sim10^9$ galaxies over 30,000 deg.$^2$ out to $z\sim2$.

\begin{figure}
\centering
\includegraphics[width=\columnwidth]{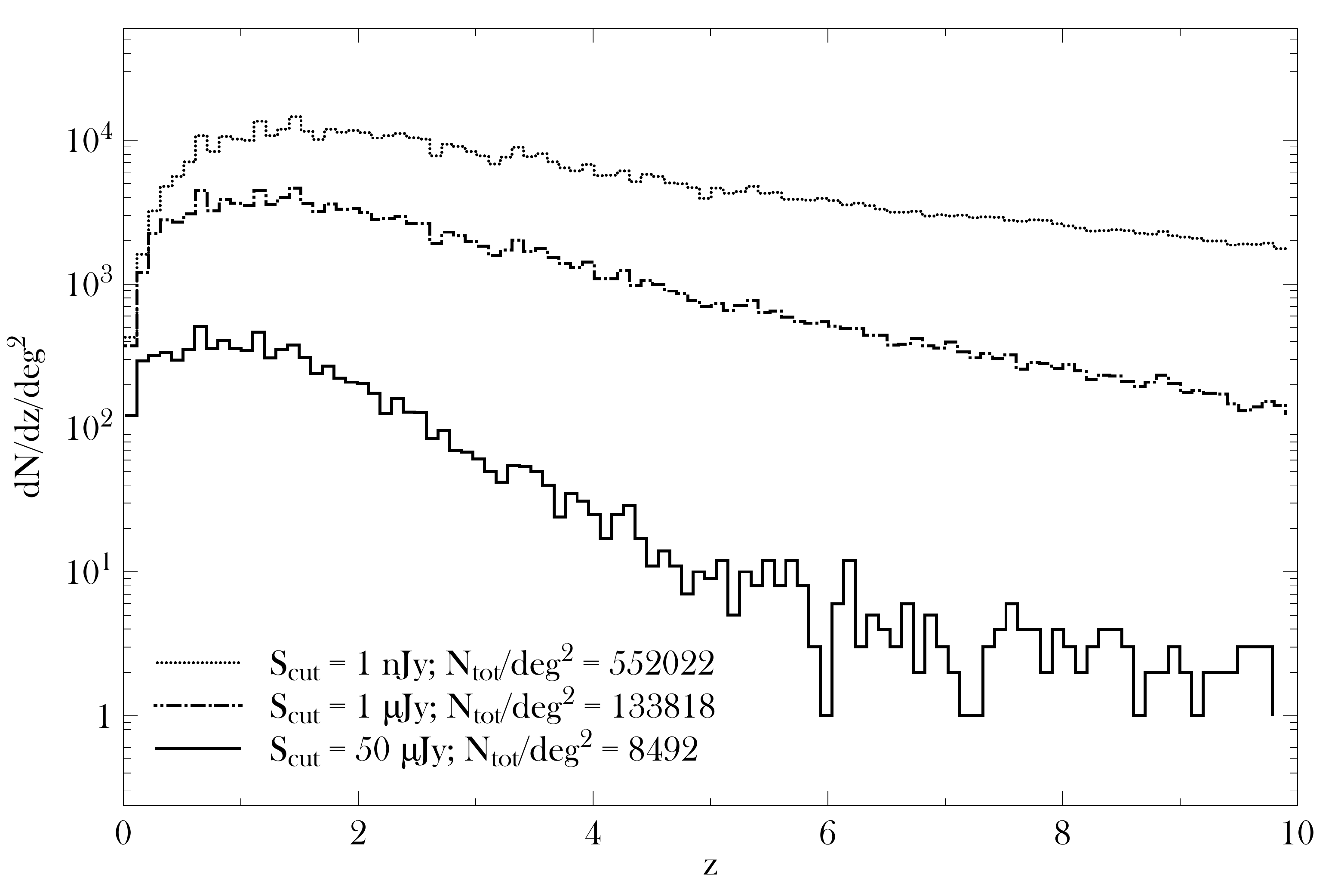}
\includegraphics[width=\columnwidth]{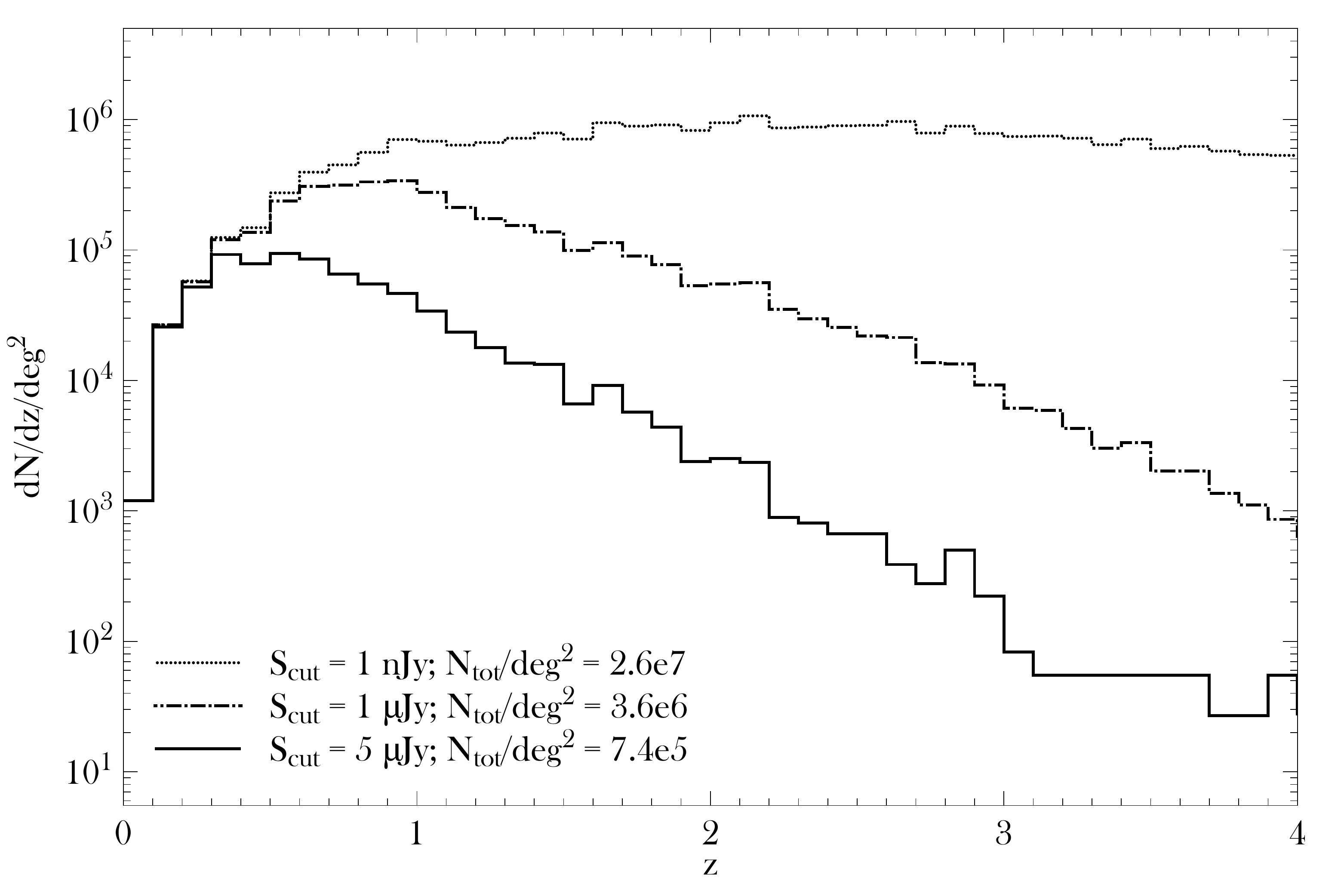}
\caption{Number of sources per sq. deg., as a function of redshift, for different flux limits. {\it Top Panel}: radio continuum surveys; {\it Bottom Panel}: H{\sc i} galaxy surveys. In the legend is the total number of objects. Distributions taken from the $S^3$ simulation.}
\label{fig:Nz_radio}
\end{figure}

In the lower panel of Figure~\ref{fig:Nz_radio} we plot the predicted redshift distributions for H{\sc i} surveys. Again, we show three different cuts in sensitivity: (i) 5 $\mu$Jy, corresponding roughly to the planned sensitivity of the SKA2 galaxy survey, (ii) a more optimistic 1$\mu$Jy, and once again (iii) a futuristic distribution including sources down to 1$n$Jy.

\bigskip

In order to keep our results as general as possible, we define surveys in terms of full sky surveys with the above flux limits.
The two cases of $S_{\rm lim} = 1\,n {\rm Jy}$ are shown as a proof of principle of what very futuristic surveys could in principle achieve.
In addition, in order to investigate what can be done in the future (and possibly helping the planning of future instruments), we will in some cases study what instrument specifications would allow specific measurements.

\subsubsection{Bias}
It is necessary to model how biased the observed sources are in relation to the underlying structures. On large scales we assume that the two-point correlation function can be written as~\citep{Matarrese:1997}:
\begin{equation}
\xi(r,z)=b^{2}(M_{\rm eff},z)\xi_{\rm DM}(r,z) \, ,
\end{equation}
where $\xi$ is the observed galaxy correlation, $b$ is the bias, $M_{\rm eff}$ represents the effective mass of dark matter halos in which sources reside and $\xi_{\rm DM}$ is the correlation function of dark matter.
For the purposes of this paper we use the bias in the $S^3$ simulation for each galaxy population, which is computed assigning each population a dark matter halo mass. This dark matter halo mass is chosen to reflect the large-scale clustering found by observations.

\begin{figure}
\centering
\includegraphics[width=\columnwidth]{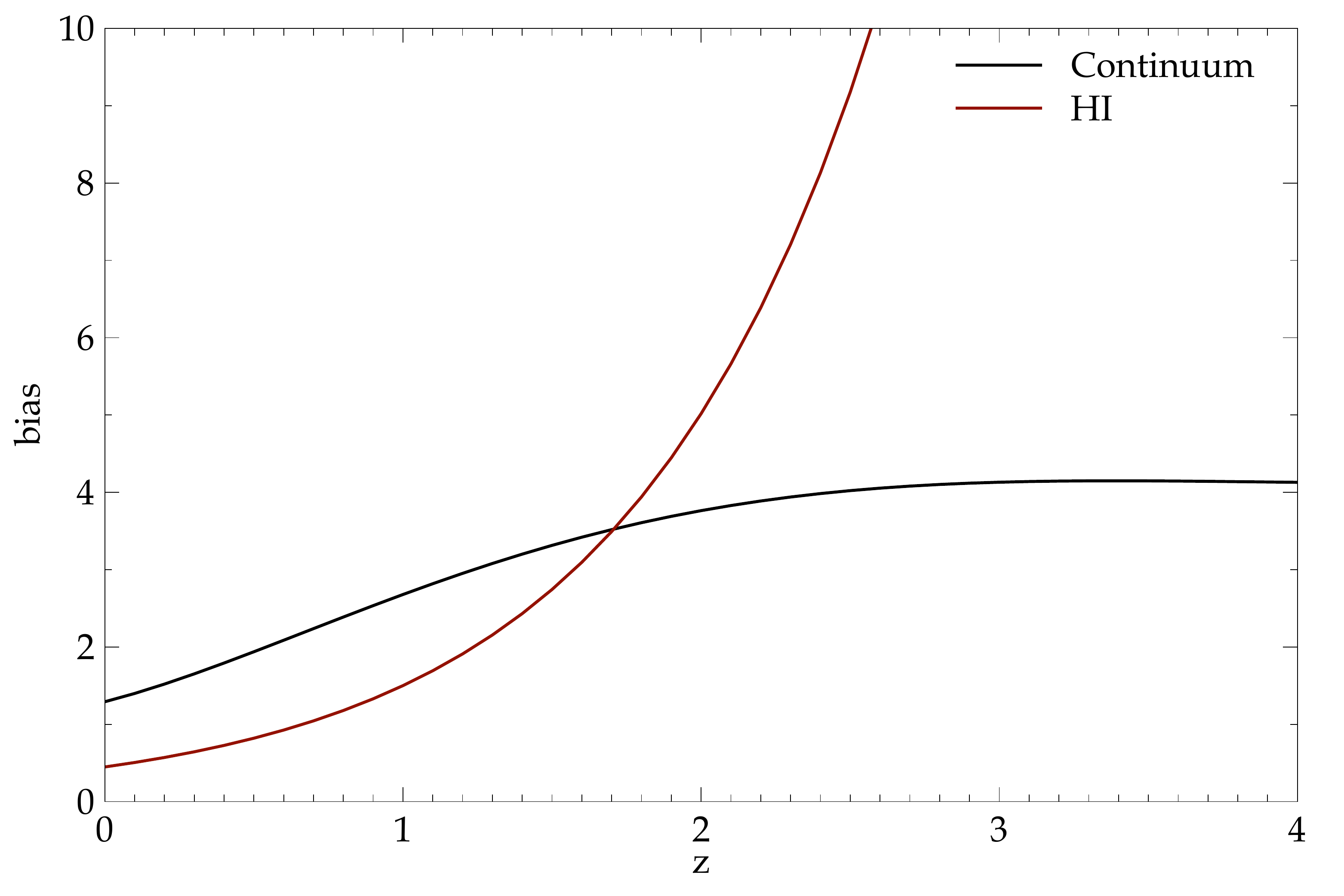}
\caption{Bias for the radio continuum (black) and H{\sc i} (red) galaxy surveys for an example flux limit of 10$\mu$Jy.}
\label{fig:bias}
\end{figure}

The $S^3$ simulation provides us with a source catalogue where sources are identified by type; each of these has a different prescription for the bias, as described in~\citet{Wilman:2008}.
We use the models of~\citealt{Raccanelli:radio} (for continuum surveys) and of~\citealt{Santos:2015} (for H{\sc i} surveys) to assign the bias to sources at different flux limits.
In Figure~\ref{fig:bias} we show an example of the bias used for continuum and H{\sc i} galaxy surveys with a flux limit of 10$\mu$Jy.

\section{Galaxy-GW Cross-Correlations}
\label{sec:ClgGW}
One of the most used observables in the analysis of galaxy surveys is the galaxy power spectrum (or correlation functions). It is possible to extract a trove of information from radial and angular correlations (see e.g.~\citealt{Raccanelli:growth, SKA:Raccanelli, Raccanelli:2015GR}), both by auto-correlating galaxies and by cross-correlating galaxies from different galaxy surveys (see e.g.~\citealt{SKA:Bacon}) or with other observables, such as CMB temperature maps (see e.g.~\citealt{Bertacca:ISW, Raccanelli:ISWfNL}). A summary of cosmological measurements from auto- and cross- correlations using radio continuum surveys can be found in~\cite{Raccanelli:radio}.

In this Section we focus on cross-correlations of radio galaxy with GW maps, both by correlating samples in the same redshift range to investigate properties of binary black hole (BBH) mergers, and by looking at radial cross-correlations in different redshift bins, in order to detect the lensing of GWs by foreground galaxies, and use this measurement to constrain models that explain the cosmic acceleration. 

In Section~\ref{sec:ccf} we briefly summarize the methodology for determining the nature of BBH progenitors, presented in~\citealt{Raccanelli:2016PBH}, and present updated forecasts.

Radial cross-correlations can be used to test models of dark energy and modified gravity, as first suggested by~\cite{Cutler:2009} and then further studied in~\cite{Camera:2013}; in Section~\ref{sec:cosmag} we will analyze this technique in the context of radio surveys correlated with BBH mergers.

\subsection{Angular cross-correlations}
\label{sec:ccf}
The angular cross-correlation of galaxy catalogs with GW maps is a natural way to investigate properties of the progenitors of compact objects whose mergers give rise to the GWs detected by laser interferometers, such as the hypothesis that BBH trace matter inhomogeneities~\citep{Namikawa:2016}, or to constrain the distance-redshift relation~\citep{Oguri:2016}.

In a similar way,~\citealt{Raccanelli:2016PBH} recently suggested that the cross-correlation of Star Forming Galaxies (SFG) with GW maps can constrain the cosmological scenario in which the Dark Matter is comprised of Primordial Black Holes (PBHs).

Primordial Black Holes were first studied in~\cite{Zeldovich:1967, Hawking:1971, Carr:1974}, and a vast literature on the topic was then produced.
The possibility that they could make up the dark matter was first investigated in~\cite{Garcia-Bellido:1996, Nakamura:1997}.
However, these models were predicting PBHs of relatively small masses ($\sim 1 M_{\sun}$), that were subsequently ruled out by microlensing experiments~\citep{Wyrzykowski:2011}.

Recently, the LIGO detection of the merger of binary black holes of larger masses ($\sim 20-30 M_{\sun}$) suggested that these objects might be quite common.
Interestingly enough, the combination of observational constraints from microlensing and disruption of wide binaries~\citep{Wyrzykowski:2011, Monroy-Rodriguez:2014} rule out the possibility of PBHs as DM for BHs with masses below $\sim 10 M_{\sun}$ and above $\sim 80 M_{\sun}$, but leave open the window in between\footnote{
CMB measurements have tentatively closed this window~\citep{Ricotti:2007}, but the result is based on several untested assumptions and until a more detailed modeling is used, one should associate a large uncertainty with this limit. Hence, we defer this issue to further studies.}.

Therefore,~\cite{Bird:2016} suggested that PBHs of $30 M_{\sun}$ comprise the dark matter; other suggestions including models with a wider mass distribution have been then studied in~\cite{Clesse:2016, Sasaki:2016}; for a recent more general review on PBHs as DM, see~\cite{Carr:2016}.
In addition, calculations of the formation of binary systems of such objects and their merger rate overlaps with estimates based on the first aLIGO observations.
Given the ongoing difficulty in detecting weakly-interacting massive particles, this model recently attracted a lot of attention and it is therefore important to develop methods to test it, and more generally determine the nature of BBH progenitors.

In this work we follow the same formalism of~\cite{Raccanelli:2016PBH}, but using an updated (including the most recent LIGO data) merger rate of 1-10 Gpc$^{-3}$ yr$^{-1}$, and a wider redshift range, with $z_{\rm max}^{\rm H{\sc i}}=3$ and $z_{\rm max}^{\rm cont}=5$, for H{\sc i} and continuum radio surveys, respectively.

We use number count measurements in order to measure the correlation between the host halos of BBH mergers and galaxies.
We consider angular power spectra $C_\ell$, that can be calculated from the underlying 3D matter power spectrum by using 
(see e.g.~\citealp{Raccanelli:2008, Pullen:2012}):
\begin{align}
\label{eq:ClXY}
C_{\ell}^{XY}(z,z') &= \left< a^X_{\ell m}(z) a^{Y\,^*}_{\ell m}(z') \right> \vs
&= r \int \frac{4\pi dk}{k} \Delta^2(k) W_{\ell}^X(k, z) W_{\ell}^Y(k, z') \, ,
\end{align}
where $W_{\ell}^{\{X,Y\}}$ are the source distribution window functions 
for the different observables (here $X$ and $Y$ stand for galaxies and GWs), $\Delta^2(k)$ 
is the dimensionless matter power spectrum today, and $r$ is a 
cross-correlation coefficient.

The window function for the number count distributions can be written as 
(see e.g.~\citealt{Cabre:2007, Raccanelli:2008}):
\begin{equation}
\label{eq:flg}
W_{\ell}^X(k) = \int \frac{d N_X(z)}{dz} b_X(z) j_{\ell}[k\chi(z)] dz \, ,
\end{equation}
where $d N_X(z)/dz$ is the redshift distribution of the species X;
$b_X(z)$ is the bias that relates the observed correlation function to the 
underlying matter distribution;
$j_{\ell}(x)$ is the spherical Bessel function of order $\ell$, and $\chi(z)$ is the comoving distance.

For our galaxy catalog we assume the redshift distributions of Section~\ref{sec:radiosurveys}; as for GW events, their number can be estimated by:
\begin{equation}
\label{eq:ngw}
\frac{d N_{GW}(z)}{dz} \approx \mathcal{R}(z) \tau_{\rm obs} \frac{4\pi\chi^2(z)}{(1+z)H(z)} \, ,
\end{equation}
where $\mathcal{R}(z)$ is the redshift-dependent merger rate, 
$\tau_{\rm obs}$ is the observation time and $H(z)$ is the Hubble parameter.
The errors in the auto- and cross-correlations are given by (see e.g.~\citealt{Cabre:2007}):
\begin{equation}
\label{eq:err-clgg}
\sigma_{C_{\ell}^{\rm X\,X}} = \hksqrt{\frac{2\left(C_{\ell}^{\rm X\,X}+ \frac{1}{\bar{n}_X}\right)^2}{(2\ell+1)f_{\rm sky}}} \, ,
\end{equation}
and:
\begin{equation}
\label{eq:err-clgt}
\sigma_{C_{\ell}^{\rm X\,Y}} = \hksqrt{\frac{\left(C_{\ell}^{\rm X\,Y}\right)^2 
+ \left[ \left( C_{\ell}^{XX} + \frac{1}{\bar{n}_X} \right) \left(C_{\ell}^{\rm 
Y \, Y}+ \frac{1}{\bar{n}_{\rm Y}} \right)\right]}{(2\ell+1)f_{\rm sky}}} \, ,
\end{equation}
where $f_{\rm sky}$ is the fraction of the sky observed and $\bar{n}_{\rm X}$ is the average number of sources per steradian, in the bin considered, of the species $X$.

A key element to determine the nature of the progenitors of BH-BH mergers is represented by the value of the halo bias of the mergers' hosts. 
While we expect that mergers of objects at the endpoint of stellar evolution will be hosted by galaxies that contain the majority of stars, and therefore be hosted in halos of $\sim 10^{11-12} M_\odot$, almost all mergers of PBH binaries would happen in halos of $< 10^{6} M_\odot$, as shown in~\cite{Bird:2016}.
Crucially, these two types of halos will have very different values of the bias. In particular, we assume that galaxies that host stellar
GW binaries have similar properties of the SFG galaxy sample. Hence, we assume that $b^{\rm Stellar}_{GW} = b_{\rm SFG}$, taking its redshift dependence as in~\cite{Ferramacho:2014}.
On the other hand, the bias of the small halos that host most of the PBH mergers is expected to be $\lesssim 0.5$, roughly constant with redshift, within our considered range~\citep{Mo:1995}.
Hence, considering that the bias of SFGs is predicted to be $b(z) > 1.4$, we set, for the threshold granting a detection, $\Delta_b = b_{\rm SFG} - b_{\rm GW} \gtrsim 1$; this value should in reality increase with redshift, making our choice conservative.

It is worth noting that this scenario can be also tested by using the eccentricity of the binaries' orbits before merging~\citep{Cholis:2016}; other ways to constrain properties of BBH mergers can be found in e.g.~\cite{Kushnir:2016, Stone:2016}.

To predict the precision with which one can test models for the nature of the BBH progenitors, we consider the effective correlation amplitude, $A_c\equiv r \times b_{GW}$, where $r$ is the cross-correlation coefficient of Equation~\eqref{eq:ClXY}. 
The cross-correlation coefficient $r$ parameterizes the extent to which two biased tracers of the matter field 
are correlated~\citep{Tegmark:1998}.
This factor accounts for the fact that the two sources are not necessarily correlated, and its value ranges from 0 (totally uncorrelated) to 1 (perfectly correlated).
In this context, an example could be the following: if sources of GW are astrophysical, for example coming from globular clusters, they would be originated in SFGs. However, if for any dynamical process they are ejected out of their host galaxy, then the correlation coefficient would be equal to the fraction of remaining merging binaries.
This effect is not important unless very high angular-resolution is achievable.
However, given that this number is also completely degenerate with the other amplitudes, we will constrain the combined quantity $A_c$ defined above.

Given the specifications of the proposed future surveys, we forecast the precision in our measurements using the Fisher matrix formalism~\citep{Fisher:1935, Tegmark:1997}:
\begin{equation}
\label{eq:Fisher}
F_{\alpha\beta} = \sum_{\ell} \frac{\partial C_\ell}{\partial 
\vartheta_\alpha}
\frac{\partial C_\ell}{\partial
\vartheta_\beta} {\sigma_{C_\ell}^{-2}} \, , 
\end{equation}
where $\vartheta_{\alpha, \beta}$ are the parameters one wants to measure, the derivatives of the power spectra $C_\ell$ are evaluated at fiducial values $\bar \vartheta_{\alpha}$ and $\sigma_{C_\ell}$ are measurement errors in the power spectra.

Given that an important source of uncertainty is given by the galaxy bias, we compute a $2\times2$ Fisher matrix for the 
parameters $\{A_c, b_g\}$, using a prior on the galaxy bias corresponding 
to a precision of 1\% in its measurement; measurements of the bias can be obtained either by fitting the amplitude of the auto-correlation of galaxy bias, or by other probes such as measurements of the bispectrum~\citep{BOSS:bispectrum} or the cross-correlation with CMB-lensing (see e.g.~\citealt{Vallinotto:2011}). Moreover, we assume the bias to be scale-independent and limit our analysis to linear scales.

We consider four different configurations of GW detectors, all assumed to observe the full sky.
~\cite{Raccanelli:2016PBH} showed that the minimum angular scale to which the GW events can be localized plays an important role in improving the constraints on the galaxy-GW cross-correlation, as does the maximum redshift observable.
An accurate determination of the value of $\ell_{\rm max}(=180^{\circ}/\theta)$ for all experiments and events is beyond the scope of this paper. Therefore, based on e.g.~\cite{LIGO-net, Namikawa:2015}, we choose the specifications as shown in Table~\ref{tab:gwdetectors}.

\begin{table}
\begin{tabular}{|l|c|c|}
\hline
Experiment & $\ell_{\rm max}$ & $z_{\rm max}$ \\ \hline
aLIGO + VIRGO & 20 & 0.75 \\
LIGO-net & 50 & 1.0 \\
Einstein Telescope & 100 & 5 (2) \\
Einstein Telescope binned & 100 & 5 (2) \\
\hline
\end{tabular}
\caption{Specifications of GW detectors used in this paper. For cross-correlations with the Einstein Telescope we use $z_{\rm max}=2$ for H{\sc i} and $z_{\rm max}=5$ for continuum radio surveys.
}
\label{tab:gwdetectors}
\end{table}
When correlating GW maps from the Einstein Telescope (ET), we use $z_{\rm max}=2$ for H{\sc i} surveys and $z_{\rm max}=5$ for continuum surveys; for the continuum case we assume we can bin sources into bins of width $\Delta_z=1$, at the price of discarding 90\% of the sources (see~\citealt{Kovetz:2016} for details).
Considering the difficulty in determining a very precise redshift of GW events at high redshift, we take the safe assumption of having bins of $\Delta_z=1$ also for the H{\sc i} case.

\begin{figure}
\centering
\includegraphics[width=\columnwidth]{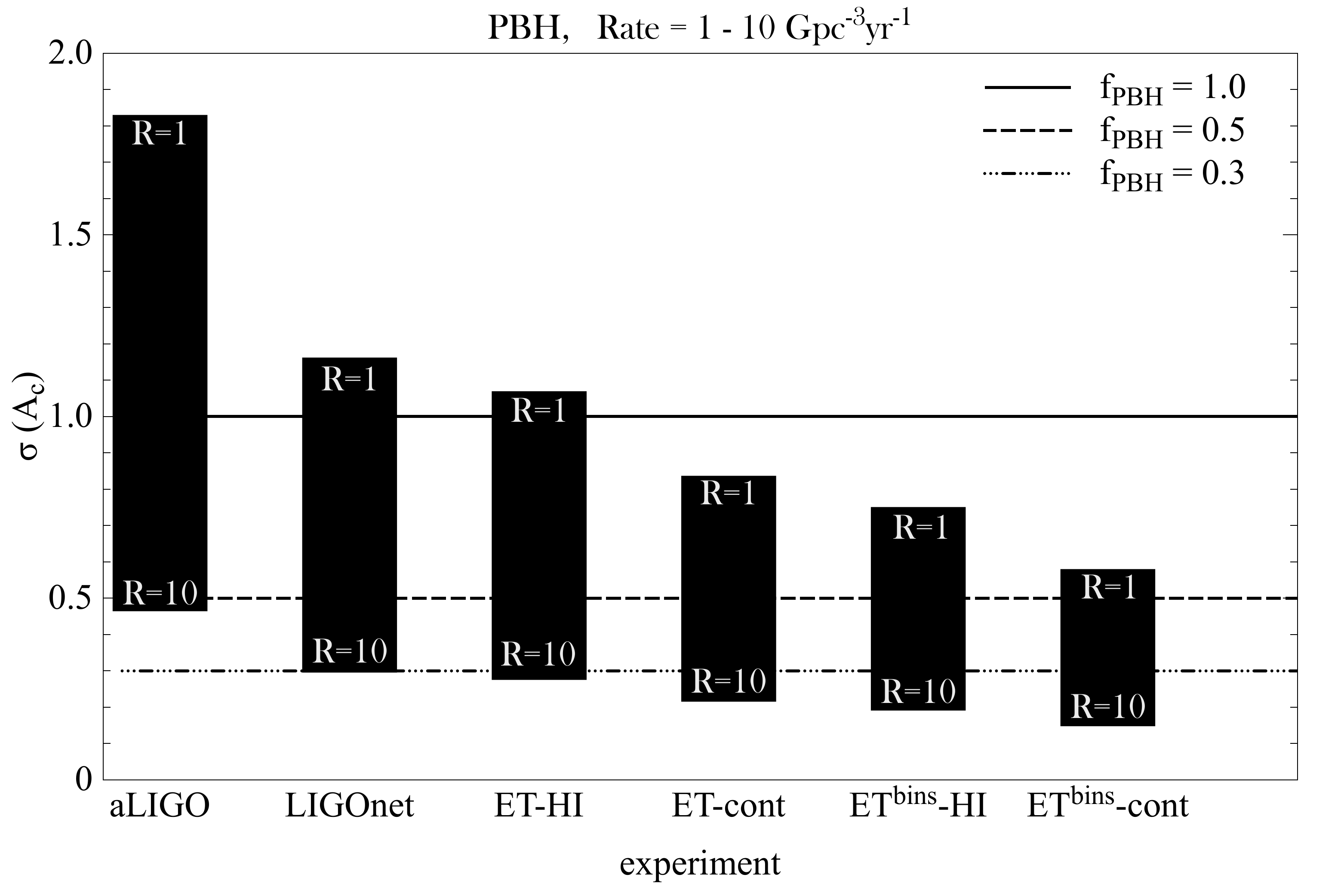}
\caption{Forecast errors on the cross-correlation amplitude, $A_c$, for different fiducial experiment sets and varying merger rates. 
Each column corresponds to a GW detector experiment, for merger rates from 1 to 10 Gpc$^{-3}$yr$^{-1}$.
The horizontal lines show the expected difference in the cross-correlation between primordial and stellar binary progenitors, for three different values of the percentage of DM made of PBHs.
}
\label{fig:pbh_exp}
\end{figure}

In Figure~\ref{fig:pbh_exp} we show our predicted constraints. We show forecasts for the different GW interferometers setups, after 5 years of collecting data, correlated with radio surveys. Vertical bars show the constraints when varying the merger rate $\mathcal{R}\in[1,10]$. Horizontal lines show the required precision in measurements of $A_c$ to test the cases where the fraction of DM made of PBHs, $f_{\rm PBH}$, is $<1$.
In the cases of aLIGO and the expanded LIGO-net, correlating with continuum or H{\sc i} surveys will not make a difference. When using ET data, on the other hand, the maximum redshift used for galaxy surveys will make a difference, thus we show separate results for the correlations GW-radio galaxy for H{\sc i} and continuum.

As one can see, already in the case of aLIGO, if the merger rate will be toward the largest limits currently allowed by observations, one should be able to detect the signature of PBHs as DM within a few years. In the more futuristic cases of correlations of ET maps with binned continuum radio surveys, we can expect to have a few $\sigma$ detection in the case of $f_{\rm PBH}=1$, or detect the effects of a small fraction of DM consisting of PBHs.

\subsection{Cosmic Magnification}
\label{sec:cosmag}
Gravitational lensing causes light rays to be deflected by large scale structures along the line of sight, introducing distortions in the observed images of distant sources~\citep{Turner:1984}.
Even though surface brightness is conserved, the sources behind a lens are magnified in size, hence inducing an increase in the total observed luminosity of a source. Therefore, sources that are just below the flux threshold of a given survey will be magnified and become detectable, and so the observed number density of sources is increased.
On the other hand, lensing causes the stretching of the observed field of view, leading to a dilution of the number density.
The net result of these two competing effects depends on the slope of the source number count as a function of the sources luminosity, and the effect is called magnification bias.
Observationally we can detect the effects of magnification by cross-correlating two galaxy surveys with disjoint redshift distributions, as the {\it observed} number counts can be modified as explained above, and the intrinsic galaxy clustering is negligible for long radial correlations.
The implications of this effect for the observed galaxy angular correlation function were investigated in e.g.~\cite{Villumsen:1997, Kaiser:1998}.
Cosmic magnification has been suggested as a probe for cosmology by~\cite{Matsubara:2000} and has been subsequently studied in a variety of works (see e.g.~\citealt{Hui:2007, Loverde:2008}).
Cosmic magnification was first detected by~\cite{Scranton:2005}, who cross-correlated foreground SDSS LRGs with a background of SDSS quasars, and then e.g. in the Canada-France-Hawaii-Telescope Legacy Survey~\citep{Hildebrandt:2009}, and Herschel~\citep{Wang:2011}.

The effect of cosmic magnification has been recently investigated in even more detail in the context of large-scale contributions to the {\it observed} galaxy correlation (see e.g.~\citealp{Raccanelli:radial, Montanari:2015, Cardona:2016, Raccanelli:2015GR}).
Lensing of GWs, in the case of NS-NS mergers, where the source have optical counterparts, has been considered as a probe to constrain models for cosmic acceleration~\citep{Cutler:2009, Camera:2013}, showing very promising results for future GW detectors.

Here we investigate the possibility to cross-correlate foreground galaxies that act as lenses with a background of GWs. By taking redshift bins sufficiently wide and separated, there will be no need to have a precise estimation of the redshift of BH-BH mergers.
Such measurement will allow additional and independent tests of general relativity and dark energy models, in a way that is independent from current tests using galaxy surveys alone or in combination with the CMB, hence providing a very interesting measurement.

Magnification bias introduces a correction to the {\it observed} galaxy overdensity, which now becomes the sum of the intrinsic galaxy overdensity and the magnification bias correction, $\delta_{\rm obs} = \delta_g + \delta_\mu$.
The magnification bias correction is given by:
\begin{eqnarray}
\label{deltamu}
\delta_\mu = (5s - 2) \kappa \, ,
\end{eqnarray}
where $\kappa$ is the lensing convergence:
\begin{eqnarray}
\kappa(\chi,\thetaB) = \int_0^\chi d\chi' {\chi'(\chi-\chi') \over \chi} \nabla^2_\perp (\Phi+\Psi) \, ;
\end{eqnarray}
here $\Phi, \Psi$ are the gravitational potentials and
$\nabla^2_\perp$ is the 2D Laplacian in the transverse direction. 
The magnification bias $s$ is defined as:
\begin{eqnarray}
s = {d {\,\rm log} N_{|_{< M}} \over d M}\bigg|_{M_{\rm lim}} \, ,
\end{eqnarray}
where $M_{\rm lim}$ is the magnitude (or flux) limit of the survey and $N_{|_{< M}}$ is the number count for galaxies brighter than a magnitude (or flux) $M$.

Thus, cosmic magnification changes the number count of sources detected at a given redshift and fixed magnitude limit as:
\begin{equation}
\label{eq:cosmag}
n^{\rm obs}(z) = n_g(z) [1 + (5s-2) \kappa] \, ,
\end{equation}
where $n^{\rm obs}, n_g$ are the observed and intrinsic number of sources, respectively, $s$ is the magnification bias and $\kappa$ is the convergence.

We will make use again of the same formalism of Equation~\eqref{eq:ClXY}, but we will correlate bins at different redshifts, and compute the uncertainty using Equation~\eqref{eq:err-clgt}. We use the predicted values of magnification bias as in~\cite{Wilman:2008, Camera:2014}, for continuum and H{\sc i}, respectively.


\subsubsection{Cosmological model constraints}
\label{sec:cosmodel}
While in principle cosmic magnification and galaxy clustering depend (and so can be used to test on) several cosmological parameters, here we focus on tests of models that explain cosmic acceleration.
To explain the accelerated expansion of the universe, it is useful to look at two possibilities, which can be easily described by looking at possible modifications of Einstein's field equations. One can introduce a dark energy component and so modify the right-hand side of the Einstein's equations, or the geometric (left) side of them:
\begin{align}
\label{eq:EE}
G_{\mu\nu}=& \, T_{\mu\nu}+T_{\mu\nu}^{\rm de}\,, \,\,\,\,\,\,\, \\
G_{\mu\nu}+G_{\mu\nu}^{\rm MG}=& \, T_{\mu\nu}\,.
\label{eq:EE2}
\end{align}
This corresponds, in the first case, to the introduction of some contribution to the energy-momentum tensor, usually in the form of a scalar field, and such a component is called dark energy.
The second alternative is based on modifications to General Relativity that lead to a modification of gravity on the largest scales and thus to acceleration, a scenario often called dark gravity or modified gravity.
Differentiating between the dark energy and the modified gravity scenarios is one of the main challenges cosmologists are facing today. \\
%


\noindent\textbf{Dynamical Dark Energy}
\label{sec:de}

\noindent Assuming the validity of General Relativity at all scales, in order to explain cosmic acceleration it is necessary implement the modifications of Equation~\eqref{eq:EE}; the simplest of such modifications is the introduction of a cosmological constant, as first suggested by~\cite{Zeldovich:1968}, that can be interpreted as vacuum energy.
A more general model, sometimes called dynamical dark energy, generalizes the cosmological constant ones by considering the time evolution of its equation of state, $w = p/\varrho$, where $p$ and $\varrho$ are the pressure and energy density of the dark energy fluid; for a cosmological constant, $w=-1$.
In this work we adopt the widely used parameterization~\citep{Linder:2003}:
\begin{align}
w(a) = w_0 + w_a (1-a) \, .
\end{align}
In this Section, we predict the precision in measurements of the pair $\{w_0, w_a\}$ from the radial correlation of (foreground) radio source number counts with background maps of GWs. \\


\noindent\textbf{Modified Gravity}
\label{sec:dg}

\noindent An intriguing alternative to dark energy for the explanation of the accelerated expansion of the universe is the ``modified gravity" approach (see e.g.~\citealt{Durrer:2007}), which investigates possible modifications, on large-scales, of gravity. In these models, cosmic acceleration can be obtained by modifying the Einstein-Hilbert action, hence evading the need of a dark energy component in the Universe.
Referring now to Equation~\eqref{eq:EE2}, in this case we add a term to the geometric side of Einstein's equations.
Modified gravity models can mimic the $\Lambda$CDM model at the level of background expansion, but in general they predict different dynamics for the growth of cosmic structures.
Here we consider scalar metric perturbations around a FRW background for which the line element in the conformal Newtonian gauge is:
\be
\label{metric}
ds^2=-a^2(\tau)\l[\l(1+2\Psi\r)d\tau^2-\l(1-2\Phi\r) d\vec{x}^2\r] \ ,
\ee
where the gravitational potentials $\Phi$ and $\Psi$ are functions of time and space. 
We use the following parameterization to describe the relations specifying how the metric perturbations relate to each other, and
how they are sourced by the perturbations of the energy-momentum
tensor:
\begin{align}
\label{gamma}
\frac{\Phi}{\Psi}&=\eta({\bf \bar{X}}), \\
\label{parametrization-Poisson}
\Psi&=\frac{-4\pi G a^2 \mu({\bf \bar{X}}) \varrho \, \delta}{k^2} \ ,
\end{align}
where $\eta$ and $\mu$ are two functions encoding the modifications of gravity, depending in principle on time and scale, ${\bf \bar{X}}=\{z, k\}$. We will consider a simple approximation where
we assume $\mu=\eta=1$ at early times, with a transition to some other values at late times. This is natural in the existing models of modified gravity that aim to explain the late-time acceleration, where departures from GR occur at around the present day horizon scales. Also, the success in explaining CMB physics relies on GR being valid at high redshifts.
To model the time evolution of $\mu$ and $\eta$ we use the following functional forms to describe the transition from unity to the constants
$\mu_0$ and $\eta_0$:
\begin{align}
\label{eq:mg_eta}
\eta(z)=&\frac{1-\eta_0}{2}\left[1+{\rm tanh}\left(\frac{z-z_s}{\Delta z}\right)\right]+\eta_0 \, ,  \\
\mu(z)=&\frac{1-\mu_0}{2}\left[1+{\rm tanh}\left(\frac{z-z_s}{\Delta z}\right)\right]+\mu_0 \, .
\label{eq:mg_mu}
\end{align}
where $z_s$ denotes the threshold redshift where gravity starts to deviate from GR, that we fix at $z=6$, and 
$\mu_0, \eta_0$ are free parameters; following~\cite{Zhao:2010}, we fix the transition width $\Delta{z}$ to be $0.05$.
Other more complicated models including scale dependent MG have been proposed and investigated, but forecasts for a variety of models is beyond the scope of this paper.

\vspace{0.75cm}

When predicting constraints on cosmic acceleration models, we parameterize our cosmology using a vector of parameters:
\begin{equation}
\label{eq:par} 
{\bf \Theta} \equiv \{\aleph, \mbox{\footnotesize $\beth$}\} \; ,
\end{equation}
where
$\aleph \in \{w_0, \eta_0\}$, $\mbox{\footnotesize $\beth$} \in \{w_a, \mu_0\}$ are the parameters we want to measure.
We use a modified version of the {\tt CLASS}\footnote{http://class-code.net/}~\citep{CLASS} code to calculate our observables, and Equation~\eqref{eq:Fisher} to calculate the Fisher matrices using the $\Lambda$CDM+GR as a fiducial model in each case.

\begin{figure}
\centering
\includegraphics[width=\columnwidth]{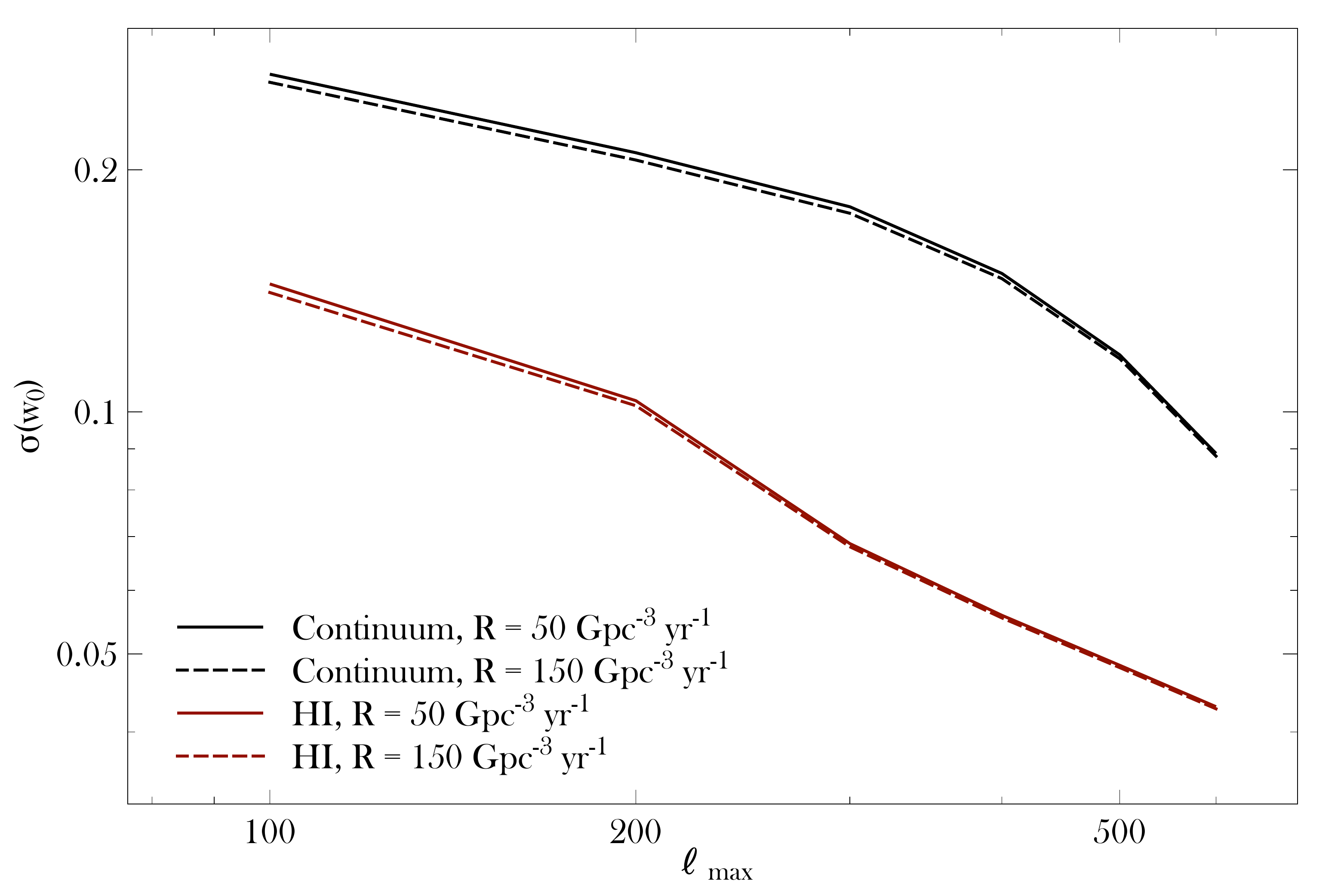}
\includegraphics[width=\columnwidth]{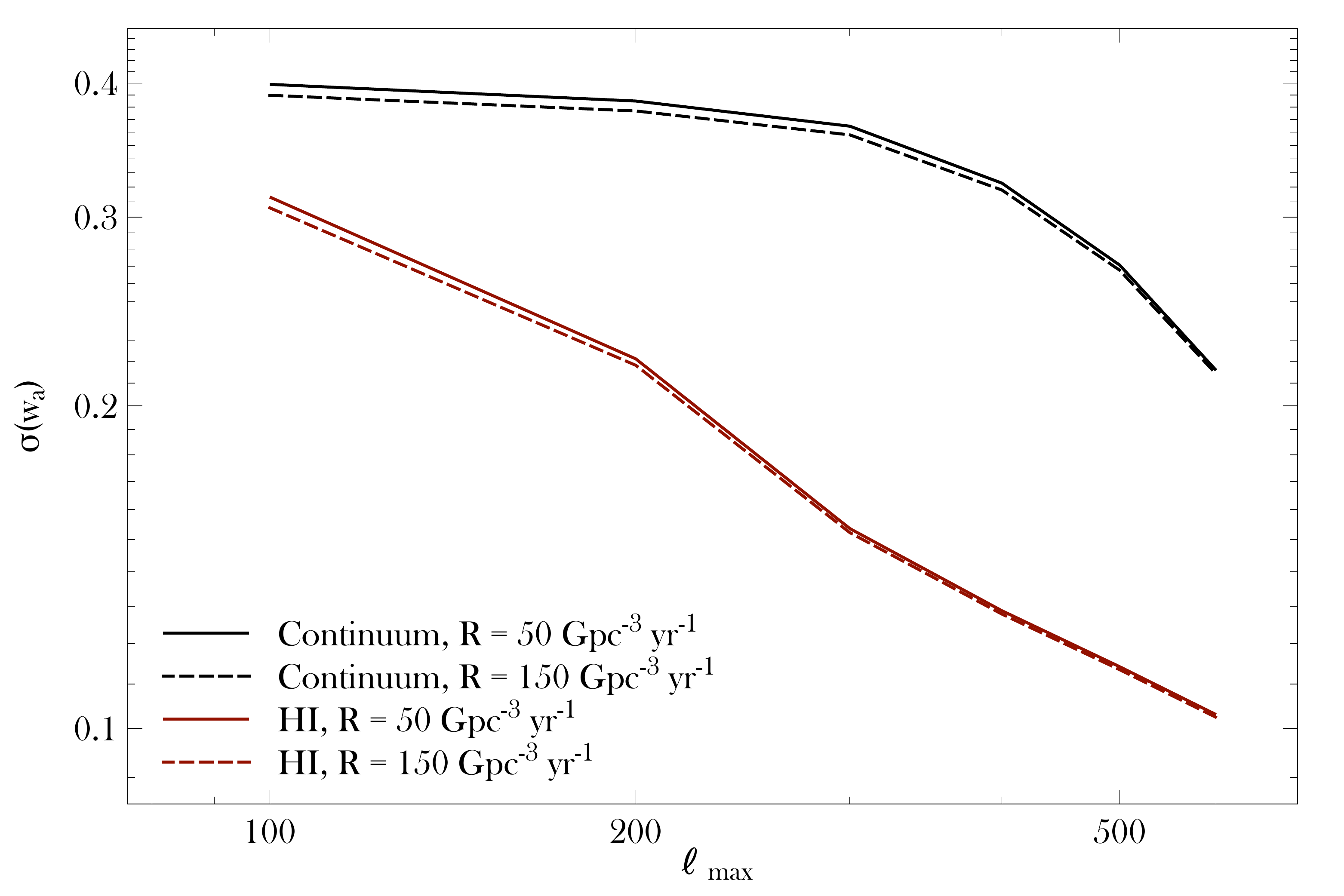}
\caption{Predicted constraints on dynamical dark energy parameters, as a function of the minimum scale used $\ell_{\rm max}$ and for two different values of the BBH merger rate $R$. Black lines show forecasts for continuum surveys, red lines for H{\sc i} spectroscopic ones.}
\label{fig:sigma_de}
\end{figure}

In Figure~\ref{fig:sigma_de} we show forecasts of the constraints on parameters for the dynamical dark energy model.
We plot constraints as a function of the minimum scale used $\ell_{\rm max}$ and for two different values of the BBH merger rate $R=50, 150$.

We use the cross-correlation of foreground galaxies that act as lenses for background GWs. The main shot noise contribution comes from the GW number counts, therefore we select a larger bin for them than the one(s) for galaxies.
In the continuum case, given that redshift information will allow only wide bins in $z$, we correlate a $0<z<1$ bin of galaxies with a $1<z<3$ bin of GWs. We show the predicted errors on dark energy parameters for this case with black lines.
In the case of the H{\sc i} surveys, high precision redshifts will be available, therefore we cross-correlate galaxies from 4 z-bins with $\Delta z = 0.25$ in the range $0<z<1$, with a background of GWs at $1<z<3$. Results for this case are shown with red lines.

Results are virtually independent on the flux limits, as results are shot noise limited from the GW bin but will not change once $n_g$ is Equation~\eqref{eq:err-clgt} is reasonably large.

As expected, the constraining power increases in the case of more bins (and so more correlations); moreover, increasing the maximum multipole used considerably improves the constraining power of this observable, as lensing is more powerful on small scales.

\begin{figure}
\centering
\includegraphics[width=\columnwidth]{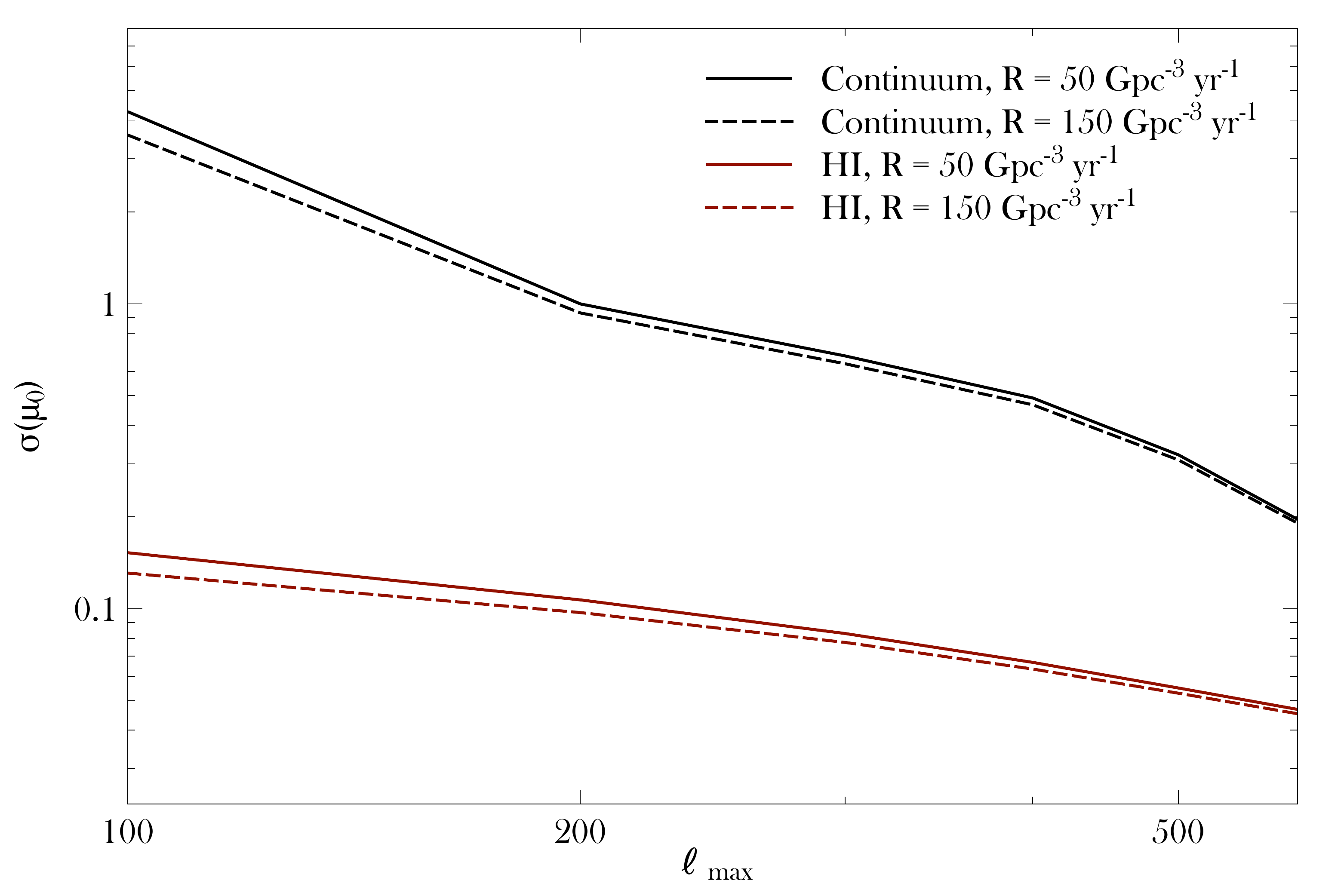}
\includegraphics[width=\columnwidth]{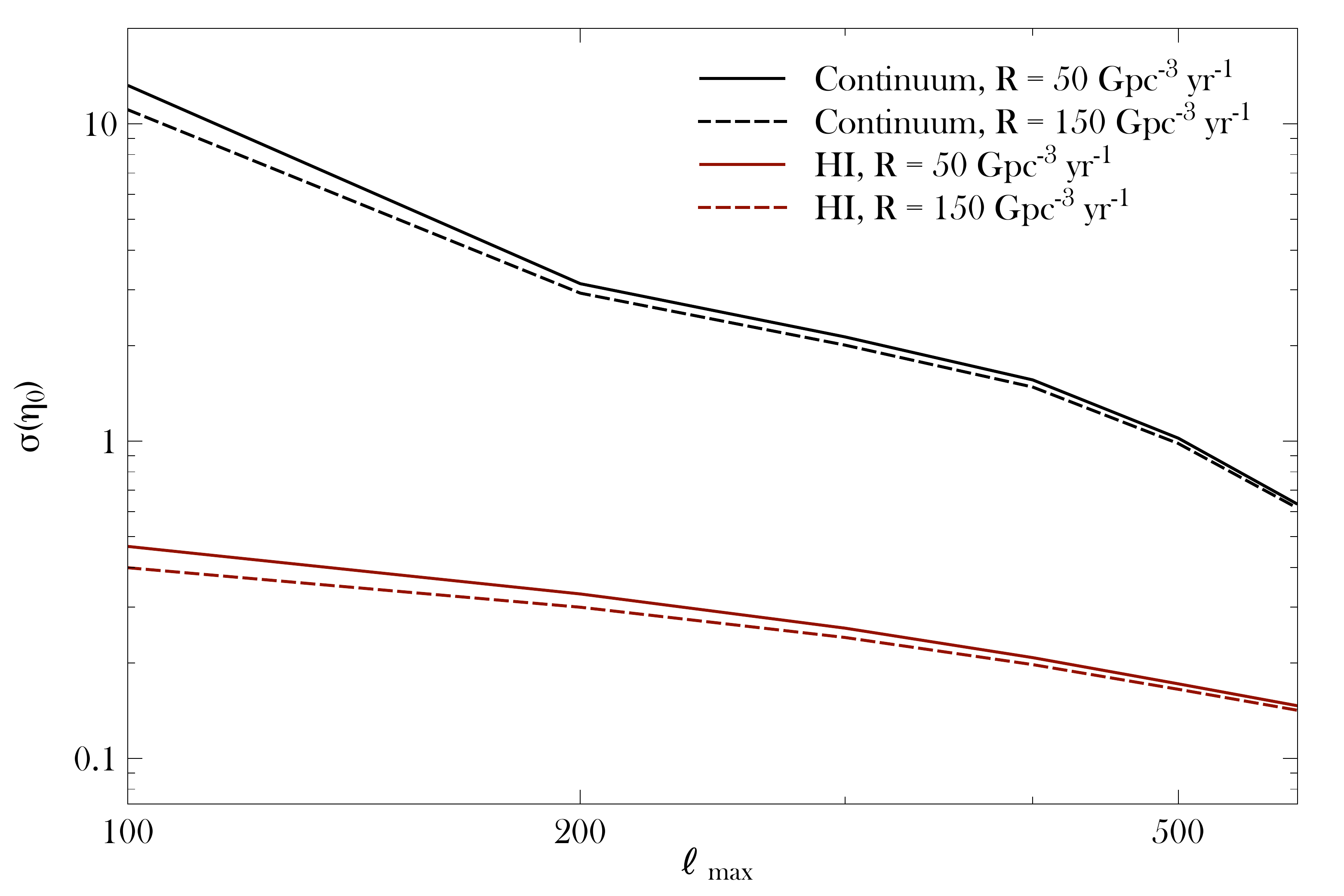}
\caption{Same as Figure~\ref{fig:sigma_de} but for modified gravity parameters.}
\label{fig:sigma_mg}
\end{figure}

In Figure~\ref{fig:sigma_mg} we show the same predictions but for the modified gravity parameters of Equation~\eqref{eq:mg_eta}--\eqref{eq:mg_mu}; results and considerations about what part of the experiment parameter space increases the constraining power are similar.

For dark energy parameters, when using H{\sc i} surveys, in the case of a conservative $\ell_{\rm max}\approx 100$, constraints will be somewhat comparable to current ones, while in the optimistic case of $\ell_{\rm max}= 600$ they will be competitive with current best constraints~\citep{Samushia:2012, Samushia:2013, Samushia:2014, BOSS:DR12} and even future ones from the combination of CMB and redshift-space distortions~\citep{SKA:Raccanelli}.

The situation is similar but even more positive for the modified gravity constraints. Our results show that, again, H{\sc i} surveys will provide better constraints than continuum ones, given the larger number of correlations measured (and small shot noise even when bins are smaller, due to the very large number density of future radio surveys).
Even when using $\ell_{\rm max}=100$, measurements of lensing of GWs from radio galaxies could provide constraints stronger than current Planck~\citep{Planck:MG} and competitive with future~\citep{SKA:Zhao} limits. However, in this case, increasing $\ell_{\rm max}$ would allow an improvement in the constraining power that is significative but not dramatic (mostly because in these models, deviations from GR are largest at large scales). Nevertheless, these results show that radial cross-correlations galaxy-GWs can be a very powerful instrument for testing deviations from Einstein's GR.

In any case, any measurements resulting from the cross-correlation galaxy-GW and lensing of GWs will represent a very important, independent and complementary check to results obtained with e.g. galaxy surveys or CMB.

It is important to note that the constraints presented here are purely indicative, and it is likely that a careful optimization could improve them.
Possible ways to increase the constraining power can come from a targeted choice of number, width and redshift center of the bins, the choice of specific galaxy populations with an optimal redshift distribution and bias; additionally, the multi-tracer technique would improve the constraints by a factor of a few.
A detailed investigation of that is however beyond the scope of this paper.

\section{Gravitational Wave Background}
\label{sec:SGWB}
A stochastic gravitational wave background (SGWB) is a key prediction of inflationary cosmology (see e.g.~\citealp{Maggiore:2000}), so that its detection is one of the main goals of astrophysics. This can be sought either via direct or indirect measurements, as the presence of the SGWB would have left imprints on different observables.

We can divide the effects of a SGWB on cosmological observables into early- and late- time effects. 
Early Universe effects include modifications to the BBN process, in particular Deuterium abundance~\citep{Pagano:2016} and imprints on the CMB: both temperature and polarization of CMB photons are affected by the SGWB, in particular the formation of a B-mode pattern in the polarization~\citep{Kamionkowski:1997, Seljak:1997, Polnarev:2008}.

Regarding late-time effects, the presence of a GW background modifies the statistics of primordial curvature perturbations and creates tidal effects that affect clustering of structures~\citep{Masui:2010, Schmidt:2012, Dai:2013, Schmidt:2014}.
Moreover, the presence of tensor perturbations causes projection effects that change the {\it observed} matter density field~\citep{Jeong:2012} and  causes a distortion of galaxy shapes~\citep{Dodelson:2003, Dodelson:2010, Schmidt:2012}.

Furthermore, GWs have an effect on local light signals propagating from closer objects; this effect could be captured by pulsar timing array observations (see e.g.~\citealp{Joshi:2013}). 

A measurement and characterization of the SGWB power spectrum will allow tests of early universe models (even though it has been seen as a possible experimental confirmation for inflation, primordial gravitational waves can also be generated in alternative models such as ekpyrotic models~\citep{Ito:2016}), and constrain fundamental parameters such as the tensor-to-scalar ratio $r$~\citep{Smith:2006, Smith:2008}.
In principle, however, the GW background could be sourced by a variety of events, including not only GWs produced during inflation, but also high-$z$ mergers of compact objects~\citep{Regimbau:2008, Regimbau:2011, Zhu:2011, Dvorkin:2016a, Dvorkin:2016b, Mandic:2016, Cholis:2016GWB}, the non-linear gravitational collapse of DM halos~\citep{Carbone:2006}, and possibly other phenomena, depending on details of the cosmological model (for a recent review, see~\citealt{Guzzetti:2016}).

Therefore, several theoretical and experimental efforts are underway for the direct and indirect measurement of the SGWB.
The main observables that are used to try to detect the SGWB are: i) direct detection of GWs; ii) B-modes of polarization in the CMB; iii) pulsar timing array experiments; iv) the anisotropy in the galaxy correlation function; v) correlations of weak gravitational lensing; vi) intensity mapping experiments probing the dark ages.

The direct detection of the SGWB will be attempted by a number of ground-based experiments that have been proposed, starting with a planned expansion of the LIGO network, with new LIGO nodes in India (IndIGO,~\citealt{INDIGO}) and Japan (KAGRA,~\citealt{KAGRA}).
Future GW detectors include the Einstein Telescope (ET,~\citealt{ET}) and the space instrument eLISA~\citep{Klein:2015}.
However, these instruments could detect the SGWB only in the case of non-single-field slow-roll inflation models, that predict a blue inflationary power-spectrum~\citep{Guzzetti:2016}.
In order to detect a scale-invariant inflationary power-spectrum, more futuristic instruments~\citep{futureGW}, such as the planned DECI-Hertz Interferometer Gravitational wave Observatory (DECIGO,~\citealt{DECIGO}) and BBO~(\citealp{BBO}), designed primarily to detect the primordial SGWB, are required~\citep{Moore:2015}.

As for the indirect detection, several ground-, balloon- and space- based CMB polarization experiments are under construction or have been proposed~\citep{SPIDER, PIXIE, CORE, CLASSCMB, PRISM, LITEBIRD, ACTPOL, POLARBEAR},
and pulsar timing array experiments are underway~\citep{PPTA, NANOGRAV, SKA:PTA, Lentati:2015}.

The strength of the SGWB is parameterized by their energy density per unit logarithmic frequency, $\Omega_{\rm GW}(f)$:
\begin{equation}
\label{definizioneomega1}
	\Omega_{\rm GW} = \frac{1}{\rho_{\rm c}}\frac{\mathrm{d}\rho_{\rm GW}}{\mathrm{d\,ln} f}.
\end{equation}
where $\rho_{\rm GW}$ is the gravitational energy density and $\rho_{\rm c} = 3H^{2}/8\pi G$ is the
critical energy density of the Universe.
Current data provide bounds on its amplitude and spectral tilt.

Current observational upper limits on $\Omega_{\rm GW}$ include (i) the constraint $\Omega_{\rm GW} \lesssim 10^{-13}$ for $10^{-17}~{\rm Hz} \lesssim f \lesssim 10^{-16} \, {\rm Hz}$ from large angular scale fluctuations in the cosmic microwave background temperature~\citep{Buonanno:2007}; (ii) the combination of cosmological nucleosynthesis and cosmic microwave background constraint gives $\Omega_{\rm GW}(f) \lesssim 10^{-5}$ for $f\gtrsim 10^{-15}$ Hz \citep{Smith:2006CMB, Pagano:2016}; (iii) the pulsar timing limit provided by EPTA, $\Omega_{\rm GW}<1.2\times 10^{-9}$ for $f=2.8\times 10^{-9}$ Hz \citep{Lentati:2015}; (iv) joint analysis of LIGO and Virgo provides an upper limit of $\Omega_{\rm GW}<5.6\times10^{-6}$ for $f\sim100$ Hz \citep{Aasi:2014}, and the most recent aLIGO analysis gave $\Omega_\text{GW} = 1.1_{-0.9}^{+2.7} \times 10^{-9}$ for $f=25$ Hz~\citep{LIGO:SGWB}; (v) limits from very long baseline interferometry radio astrometry of quasars of $\Omega_{\rm GW} \lesssim 10^{-1}$ for $10^{-17} \, {\rm Hz} \lesssim f \lesssim 10^{-9} \, {\rm Hz}$~\citep{Gwinn:1997} and $\Omega_{\rm GW} \lesssim 4\times10^{-3}$ for $ f \lesssim 10^{-9} \, {\rm Hz}$~\citep{Titov:2011}; (vi) $\Omega_{\rm GW} \lesssim 10^{-3}$ for $10^{-16} \, {\rm Hz} \lesssim f \lesssim 10^{-10} \, {\rm Hz}$ from the observed galaxy correlation function~\citep{Linder:1988}.

From CMB data, the joint analysis of Planck, BICEP2 and Keck Array data provides an upper bound of $r_{0.05}<0.09$ at $95\%$ C.L. at frequencies $f\sim 10^{-17}$ Hz~\citep{BICEP2}.

For more details on the SGWB, see the review articles~\cite{Allen:1997, Maggiore:2000, Buonanno:2007}, and more recently~\cite{Guzzetti:2016}; for a recent review on the search of gravitational waves from inflation using the CMB, see~\cite{KK}. \\

In the weak-field limit, where GWs can be described as space-time ripples propagating on a fixed background, one can write the Einstein equations in vacuum as~\citep{Guzzetti:2016}:
\begin{equation}
	\bar{G}_{\mu\nu}=\bar{R}_{\mu\nu}-\frac{1}{2}\bar{R}\bar{g}_{\mu\nu}=\langle R_{\mu\nu}^{\left(2\right)}\rangle-\frac{1}{2}\bar{g}_{\mu\nu}\langle R^{\left(2\right)}\rangle\,.
\end{equation}
where the bar $\bar{}$ indicates quantities evaluated at the background, and $\langle ...\rangle$ indicates an average over several wavelengths. 

This describes how the presence of GWs affects the background metric; following~\cite{Gravitation}, the stress-energy tensor of gravitational waves is:
\begin{equation}
\label{energygw}
	t_{\mu\nu}=\frac{1}{32\pi G}\langle \partial_{\mu}h_{ij}\partial_{\nu}h^{ij}\rangle \, ,
\end{equation}
and so the GW energy-density can be written as:
\begin{equation}
\label{densitygw}
\rho_{\rm GW}=\frac{1}{32\pi G a^{2}}\langle h'_{ij}\left(\textbf{x},\tau\right)h'^{ij}\left(\textbf{x},\tau\right)\rangle\,.
\end{equation}

\bigskip

In this paper we focus on effects of the SGWB on the LSS; a background of gravitational waves will leave an imprint in the angular correlation of galaxies on large scales, and introduce correlations of galaxy shapes. However, LSS measurements will provide weaker constraints than the CMB.
Realistically, CMB experiments will measure $r$, and if its value is not negligible (around the upper limit of current constraints), LSS could confirm or at least support those measurements.

\subsection{Cosmometry}
\label{sec:cosmometry}
As suggested by~\cite{Linder:1986, Braginsky:1990, Kaiser:1997}, a stochastic gravitational wave background causes the apparent positions of distant sources to fluctuate, with angular deflections of order the characteristic strain amplitude of the gravitational waves, and these fluctuations may be detectable with high precision astrometry.
Estimates of the upper limits obtainable on the gravitational wave spectrum $\Omega_{\rm GW}(f)$, at frequencies of order $f\sim 1 \, {\rm yr}^{-1}$ are available in literature, and they generally predict constraints comparable with the ones from pulsar timing (see e.g.~\citealt{Jaffe:2004, Book:2011}).

In this paper we extend those predictions to future radio surveys and try to understand what are the instrument requirements needed to make cosmological radio galaxy high-z astrometry, which we call ``cosmometry'', competitive with other methods and possibly have a detection of the effects of a SGWB.

\subsubsection{Effect of a SGWB on galaxy positions}
The SGWB produces an apparent angular deflection $\delta {\bf n}({\bf n},\tau)$ on the position of a source in the direction ${\bf n}$, given by~\citep{Pyne:1996}:
\be
\delta n^{i}(\tau,{\bf n}) = \frac{n^i + p^i}{2 (1 + \mathbf{p}\cdot\mathbf{n})} h_{jk}(\tau,{\bf 0}) n_j n_k - \frac{1}{2} h_{ij}(\tau,{\bf 0}) n_j \, ,
\ee
where $\mathbf{p}$ is the direction of propagation of the GW.

The apparent angular deflection caused by such a GW background is a stationary, zero-mean, Gaussian random process.
In this section we follow the methodology suggested in~\cite{Jaffe:2004} and~\cite{Book:2011}.
The total power in angular fluctuations is then:
\be
\label{eqn:2ptnn}
\left< \delta {\bf n}({\bf n},t)^2 \right> = \frac{1}{4 \pi^2} \int d\ln f \left( \frac{H_0}{f} \right)^2 \Omega_{\rm GW}(f) \, .
\ee

Specifically, a SGWB will cause apparent angular deflections which are correlated over the sky and which vary randomly with time, with a rms deflection $\delta_{\rm rms}(f)$ per unit logarithmic frequency interval of:
\be
\delta_{\rm rms}(f) \approx h_{\rm rms}(f) \approx \frac{H_0}{f} \hksqrt{\Omega_{\rm GW}(f)} \, .
\ee
By monitoring the position of sources in the sky with an angular accuracy of $\Delta \theta$, over a time T, one could detect a proper motion of order $\sim\Delta \theta/T$. For N sources, the correlated angular velocity detectable would scale as $\hksqrt{N}$. Therefore, assuming a flat GW spectrum, one should obtain an upper limit on $\Omega_{\rm GW}$ of order~\citep{Pyne:1996, Book:2011}:
\be
\label{eqn:omapprox}
\sigma_{\Omega_{\rm GW}} \approx \int_{f <T^{-1}} d \ln f \frac{\Delta \theta^2}{N T^2 H_0^2} \,\,\, .
\ee
In radio, using VLBI radio interferometry, the SKA could be able to localize sources to within $\sim 10 \mu\text{as}$~\citep{Fomalont:2004}.
With that precision, and using:
\begin{equation}
\sigma_{\Omega_{\rm GW}} \approx \frac{4 \pi f_{\rm sky}\Delta \theta^{2}}{N} \, ,
\end{equation}
it has been estimated that observing $10^6$ QSOs over a year, the SKA could set a limit of the order $\Omega_{\rm GW} \lesssim 10^{-6}$, for $f\sim 10^{-8}$Hz~\citep{Jaffe:2004}. \\

\begin{figure}
\centering
\includegraphics[width=\columnwidth]{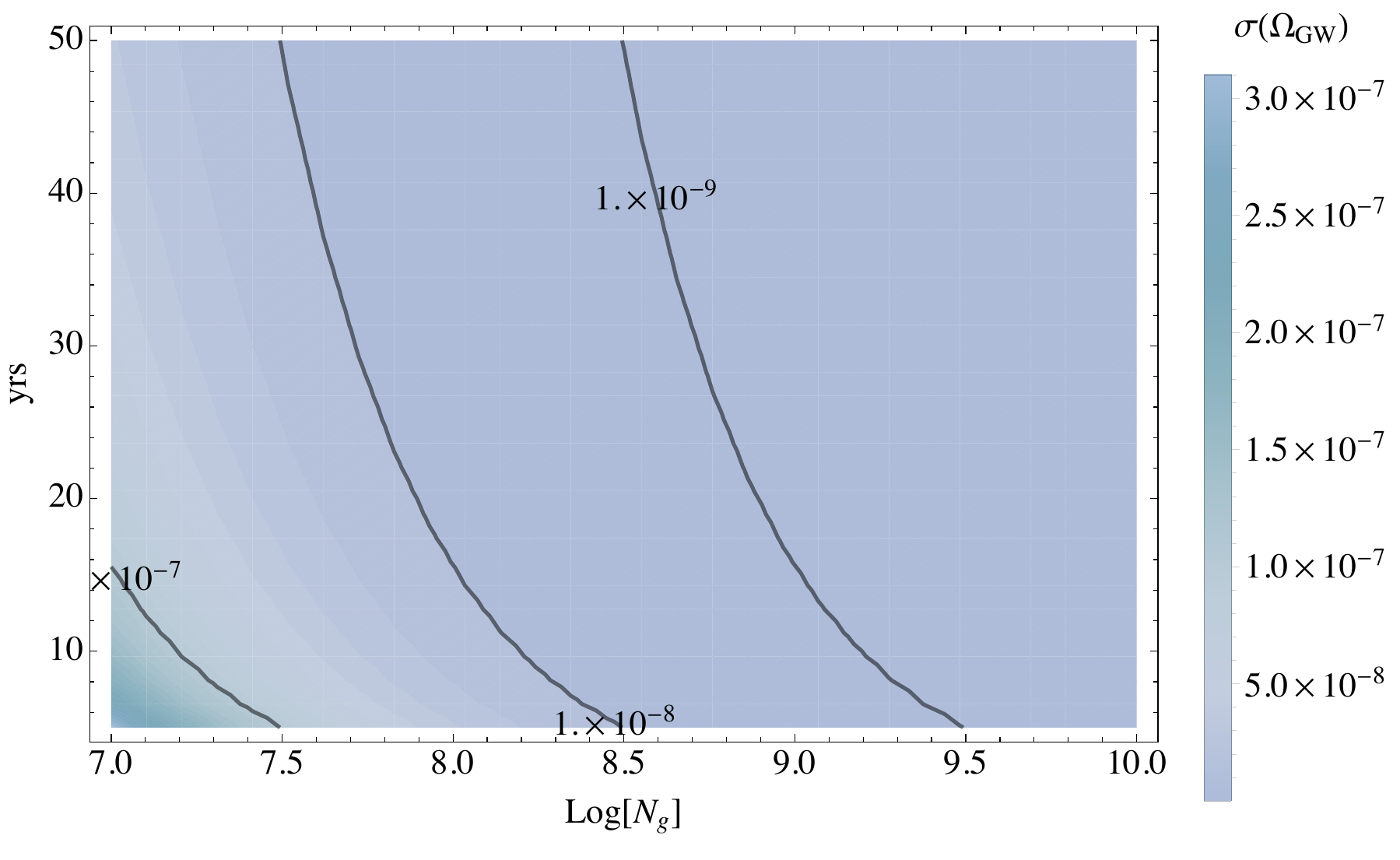}
\caption{Predicted precision in the measurements of $\Omega_{\rm GW}$ from cosmometry analyses, as a function of the number of objects and years of observation, assuming an angular resolution of 10 $\mu$arcsec.
}
\label{fig:sighcNT}
\end{figure}

\begin{figure}
\centering
\includegraphics[width=\columnwidth]{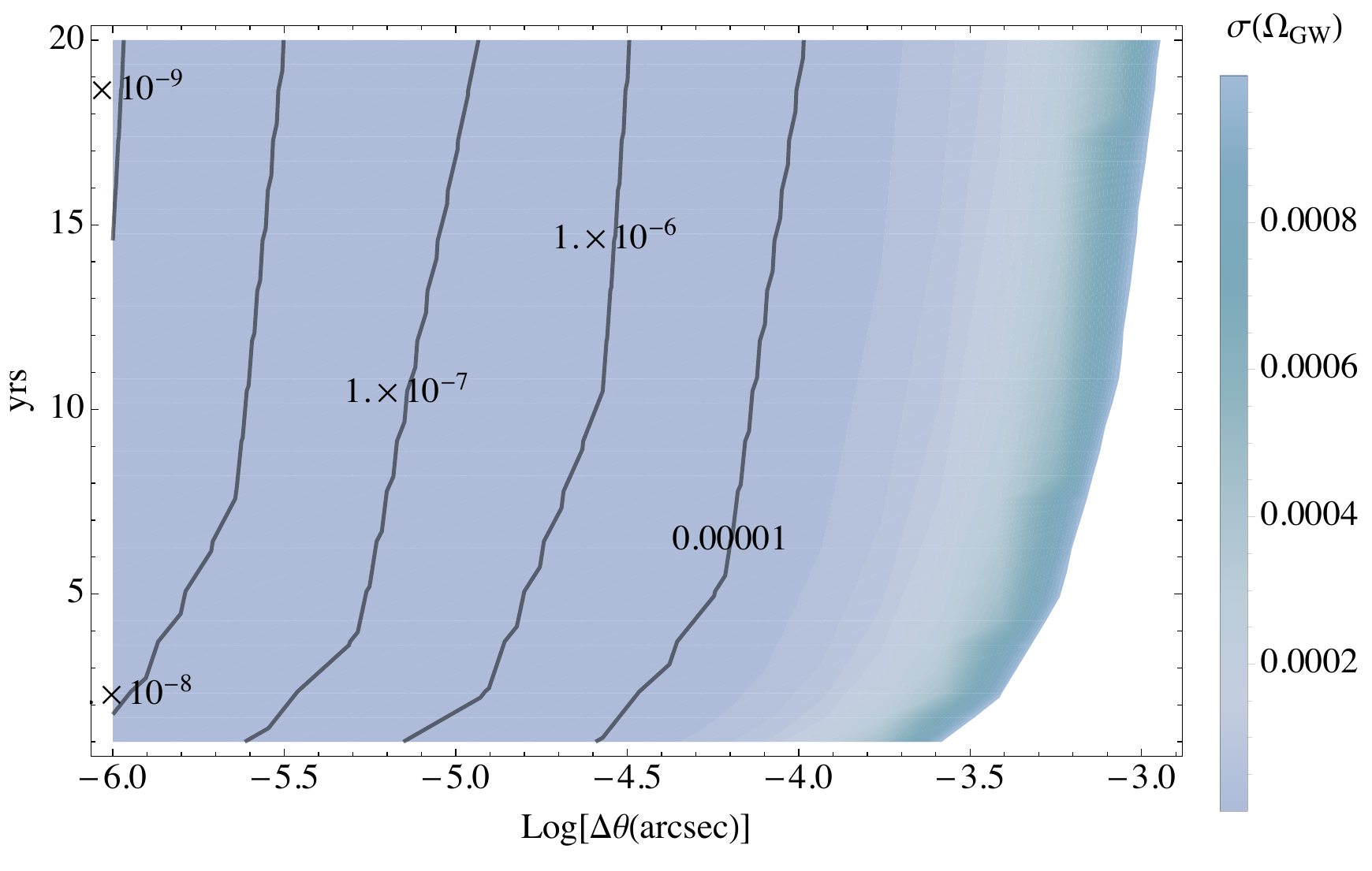}
\caption{As in Figure~\ref{fig:sighcNT} but as a function of the angular resolution and years of observation, assuming $10^7$ objects.
}
\label{fig:hctT}
\end{figure}

Here we investigate limits on $\Omega_{\rm GW}$ that will be possible to set by using cosmometry from future radio surveys.
In Figures~\ref{fig:sighcNT} and~\ref{fig:hctT} we show predicted upper limits on $\Omega_{\rm GW}$ as a function of angular resolution, observing time and number of objects observed. We focus on GWs with a period of 1 year.

We can see that, in order to detect the SGWB from changes in galaxy positions, extremely precise measurements will be needed.
Even with a very optimistic SKA in VLBI configuration, a detection of GWs from inflation is virtually out of reach.
However, configurations with baseline lengths up to $10,000$~km in length are being considered~\citep{SKA:VLBI}, therefore very precise measurements might be possible in a (relatively) near future.
In any case, limits set with this methodology are complementary to the ones set by e.g. direct detection of GWs, making results from forthcoming instruments useful and interesting.

\subsection{Cosmic Rulers}
\label{sec:rulers}
In this Section we investigate the possibility of using LSS standard rulers to detect the effects of a background of GWs; we make use of the ``Cosmic Rulers'' formalism~\citep{Schmidt:2012CR}, and investigate if future radio surveys will be able to detect the SGWB by using either the anisotropy of the 2-point galaxy correlation or galaxy ellipticities, as first suggested by~\cite{Jeong:2012, Schmidt:2012}.
We will briefly review the methodology introduced in the above series of paper, that we will use to obtain our results, and then show forecasts for measurements that will be possible to obtain with future radio surveys.
Note that, apart from a slightly different notation and some rearrangements, all the results here coincide with the ones in the cited ``Cosmic Rulers'' series. For a direct translation between the two notations, one can compare~\cite{Bertacca:2012} with~\cite{Jeong:2011GR}.

\subsubsection{Galaxy clustering}
In the context of a general relativistic description of the {\it observed} galaxy clustering (see e.g.~\citealp{Yoo:2009, Jeong:2011GR, Bertacca:2012}), a series of effects need to be considered as corrections to the standard, newtonian, formalism, which includes only scalar perturbations to the matter density field and Redshift-Space Distortions (RSD).
In a similar way, gravitational waves (tensor perturbations) affect, on the largest scales, the clustering statistics via volume and magnification effects.

Differences in the effect of tensor perturbations from scalar ones were firstly noted in~\cite{Kaiser:1997}: in particular, tensor modes redshift away, so that their contribution to the clustering of LSS tracers is dominated by contributions close to the time of emission; instead, scalar perturbations grow, and the deflection in amplified by transverse modes~\citep{Jeong:2012}.

Gravitational waves are waves in the transverse ($\partial^i h_{ij}$) and traceless ($h^i_{\,i}$) components of the metric perturbation, defined in the FRW Universe in terms of the spatial components of the metric by $g_{ij} = a^2(\delta_{ij} + 2 h_{ij})$.

We now compute the contribution of tensor perturbations to the observed galaxy clustering on large scales.
We start by considering a spatially flat FRW background, consider only tensor modes, and we write the (perturbed) conformal metric as:
\be
d\bar{s}^2 
=
-d\eta^2 + \left(\delta_{ij}+h_{ij} \right) dx^idx^j \, ,
\label{eq:metric_conf}
\ee
where $\eta$ is the conformal time and $h_{ij}$ is a metric perturbation which is 
transverse and traceless:
\be
h^i_{\; i} = 0 = (h_{ik})^{,i}.
\ee
This can be decomposed into Fourier modes of two polarization states:
\ba
h_{ij}(\mathbf{k}, \eta) = e^+_{ij}(\hat{\mathbf{k}}) h^+(\mathbf{k},\eta) + e^\times_{ij}(\hat{\mathbf{k}})
h^\times(\mathbf{k}, \eta) ,
\label{eq:hpol}
\ea
where $e^s_{ij}(\hat{\mathbf{k}})$, $s=+,\times$, are transverse 
(with respect to $\hat{\mathbf{k}}$) and 
traceless polarization tensors.
We assume both polarizations to be independent and to have equal power spectra:
\ba
\< h_{s}(\mathbf{k},\eta) h_{s'}(\mathbf{k}',\eta') \> = (2\pi)^3 \d_D(\mathbf{k}-\mathbf{k}') 
\d_{ss'} \frac14 P_T(k,\eta,\eta') ,
\label{eq:PT}
\ea
where:
\ba
P_{T}(k,\eta,\eta') = T_T(k,\eta) T_T(k,\eta') P_{T0}(k) \, ;
\label{eq:PT2}
\ea
we can write the tensor transfer function $T_T(k,\eta)$ as~\citep{Guzzetti:2016}:
\be
T_T(k,\eta) = \frac{3 j_1(k \eta)}{k\eta} \, ,
\label{eq:TT}
\ee
because tensor modes propagate as free waves after recombination.Here $P_{T0}(k)$ is the primordial
tensor power spectrum, which is defined as:
\ba
P_{T0}(k) = 2\pi^2 \Delta_T^2 k^{-3} \left(\frac{k}{k_*}\right)^{n_T} \, ,
\label{eq:PT0}
\ea
where $k_* = 0.002$, and the tensor index satisfies the inflationary consistency relation, $n_T = -r/8$.

Throughout, we will assume a tensor-to-scalar ratio of $r = 0.09$ at $k_*$ (consistent with
current upper limits), which together with our fiducial cosmology determines the amplitude $\Delta_T^2$. \\

The observed comoving number density of galaxies $n_g^{\rm obs}$ is related to 
the true comoving number density $n_g$ through (neglecting here the so-called general relativistic corrections mentioned above):
\be
n_g^{\rm obs}(\hat{\bm{x}},\zt)
=
n_g(\bm{x},\bar{z})
\left(
1+\frac{\partial \Delta x^i}{\partial \hat{x}^i}
\right).
\label{eq:cov_dens}
\ee
The factor $n_g(\mathbf{x},\bar z)$ is the true comoving number density at the point of emission, which we expand as:
\be
n_g(\mathbf{x}, \bar z) = \bar n_g(\bar z)\:[1 + \delta_g(\mathbf{x}, \bar z)] \, ,
\label{eq:ngcom}
\ee
where $\delta_g$ is the intrinsic perturbation to the comoving number density, $\bar{z}$ is the redshift that would be measured
for the source in an unperturbed universe, and is related to $z^{\rm obs}$ through:
\be
1 + z^{\rm obs} = (1+\bar z)(1+\delta z).
\ee 
Finally, $1 + \partial_i \D x^i$ is the volume distortion due to gravitational
waves, which becomes~\citep{Jeong:2011GR}:
\be
\frac{\partial \Delta x^i}{\partial \tilde{x}^i}
=
\partial_{\chit} \Delta x_\parallel
+
\frac{2\Delta x_\parallel}{\chit} - 2 \kappa \, ,
\label{eq:jacobian}
\ee
where from now on the tilde indicates {\it observed} quantities;
here $\kappa$ is the lensing convergence, defined as:
\ba
\label{eq:kappa}
\kappa = -\frac{1}{2}\partial_{\perp i}\Delta x_\perp^i \, .
\ea
In the same way as the magnification bias of Section~\ref{sec:cosmag}, lensing affects the {\it observed} galaxy density including tensor perturbations as:
\be
\Delta_{g}^{\mathfrak{t}} = \delta_{g}^{\mathfrak{t}} + \delta_{g}^{\kappa}  = \delta_{g}^{\mathfrak{t}} - \left(\delta z +\frac{1}{4}h_\parallel
- \frac{\Delta x_\parallel}{\chit} + \kappa \right) 5s \, ,
\ee
where with $\Delta_{g}^{\mathfrak{t}}$ we indicate the galaxy density including tensor and lensing effects, while
$\delta_{g}^{\mathfrak{t}}$ includes tensor effects only.

Thus, gravitational waves affect the observed density of galaxies through
a volume distortion effect, and by
perturbing their redshifts so that the {\it observed} galaxy
density $n_g^{\rm obs}$ is modified by the so-called evolution bias:
\be
\label{eq:btdef}
b_e = \frac{d\ln (a^3\bar n_g)}{d\ln a} \, ;
\ee
for more details on this term and its effects on the {\it observed} galaxy correlation function (sometimes in literature indicated as $\alpha$~\citep{Szalay:1998}), see e.g.~\cite{Raccanelli:2010wa, Jeong:2011GR, Raccanelli:2016doppler}.

Therefore, the expression of the galaxy density including the linear-order tensor contributions can be written as~\citep{Jeong:2012}:
\ba
\Delta_{g}^{\mathfrak{t}}
&=&
(b_e - 5 s )\delta z
-(2 - 5 s)\hat\kappa
-\frac{2 - 5 s}{4} h_\parallel
-
\frac{1+\zt}{2H(\zt)} h_\parallel' - \vs
&-& \frac{2 - 5 s}{2\chit}
\left[
\int_0^{\chit} d\chi h_\parallel + \frac{1+\zt}{H(\zt)}
\int_0^{\chit} d\chi h_\parallel'
\right] -\vs
&-& \frac{H(\zt)}{2} 
\frac{\partial}{\partial \zt}
\left[
\frac{1+\zt}{H(\zt)}
\right] 
\int_0^{\chit} d\chi h_\parallel' \, .
\label{eq:dgT}
\ea

In a similar fashion as for the angular correlations we used in Section~\ref{sec:ccf}, we can write the correlations between two maps $\{X,Y\}$ as~\citep{Jeong:2012}:
\ba
\label{eq:CABl}
C^{X\,Y}_\ell &=& \frac{\ell}{2\ell+1} \sum_m \mbox{Re} \< a^{X*}_{\ell m} a^Y_{\ell m} \> \\
&=& \frac1{2\pi} \frac{(\ell+2)!}{(\ell-2)!}
\int k^2 dk\: P_{T0}(k) F_\ell^X(k) F_\ell^Y(k) \, , \nonumber
\ea
where the kernel is given by:
\ba
F_\ell^X(k) = \int d\chi\:\mathcal{W}^X(\chi) T_T(k,\eta_0-\chi) \frac{j_\ell(k\chi)}{(k\chi)^2} \, ;
\label{eq:Fl}
\ea
$\mathcal{W}^X(\chi)$ is the window function $\mathcal{W}^X(\chi) = \int_{z(\chi)}^\infty \frac{dN}{d z} W^X(\chi,z) d z $, in the same fashion of Equation~\eqref{eq:flg}.
To evaluate the total tensor contribution to the angular power spectrum of galaxies, one can set
\{X, Y\}=\{g($z_i$),g($z_j$)\} to indicate the window function for samples of galaxies at redshifts $z_{\{i,j\}}$, and:
\begin{widetext}
\ba
\label{eq:Wg}
W_g^{\mathfrak{t}}(\chi) &=&- \frac{1+\zt}{2H} \d_D(\chi-\chit) + \frac{d\ln T_T}{d\eta} \bigg[ - \frac{1+\zt}{2H}  \d_D(\chi-\chit) + \frac12 \left(b_e -1 -5s + (1+\zt)\frac{dH/d\zt}{H}\right) - \frac{5s-2}{2} \left(1+\frac{1+\zt}{H\chit}\right)\bigg] -\vs
&-& \frac{15s-6}{2\chi} + \frac{5s-2}{2\chit} + \frac{5s-2}{4} \:\ell(\ell+1) \frac{\chit-\chi}{\chi\,\chit} \, .
\ea
\end{widetext}
We omitted the redshift dependence of magnification and evolution bias for convenience of notation.

Now, in order to detect the effects of tensor perturbations and so of a GW background on galaxy correlations, we need to have a measurement with errors smaller than the effect we want to detect.
Total contributions to the observed galaxy correlations from tensor perturbations are negligibly small compared to scalar ones.
The full {\it observed} galaxy 2-point angular correlation function, including all effects, can be written as~\citep{Raccanelli:2015GR}:
\begin{equation}
C_{\ell}^{ij} = 4\pi \int \frac{dk}{k} \Delta^{X}(k,z_i) \Delta^{Y}(k,z_j) \mathcal{P}(k) \; ,
\end{equation}
where the $\Delta^X$ are summed over all contributions:
\begin{equation}
\Delta_g^{\rm obs} = \Delta_g^{\delta} + \Delta_g^{\rm v} + \Delta_g^{\rm \kappa} + \Delta_g^{\Phi} + \Delta_g^{\mathfrak t} \, .
\end{equation}
The different terms account for the effects of peculiar velocity (RSD and Doppler terms, $\Delta_g^{\rm v}$), lensing ($\Delta_g^{\rm \kappa}$), gravitational potential ($\Delta_g^{\Phi}$) and tensor perturbations ($\Delta_g^{\mathfrak t}$). We can write them as follows:
\ba
\Delta_g^{\delta} &=& b \;\delta \, ;\\
\Delta_g^{\rm v} &=& \frac{1}{{\cal H}}\partial_\mathfrak{r}({\bf v}\cdot{\bf n}) 
+ \left[\frac{{\cal H}'}{{\cal H}^2}+\frac{2-5s}{\mathfrak{r}{\cal H}} + 5s- b_e\right]({\bf v}\cdot{\bf n}) + \vs
&+& \left[3{\cal H}-b_e\right]\Delta^{-1} (\nabla\cdot{\bf v}) \, ; \\
\Delta_g^{\rm \kappa} &=& -\frac{2-5s}{2}\int_0^{\mathfrak{r}}d\mathfrak{r}\frac{\mathfrak{r}-\tilde{\mathfrak{r}}}{\mathfrak{r}\tilde{\mathfrak{r}}}\Delta_2(\Phi+\Psi) \, ; \\
\Delta_g^{\Phi} &=&  [5s-2]\Phi+\Psi +\frac{\Phi'}{{\cal H}} +  \left[\frac{{\cal H}'}{{\cal H}^2}+\frac{2-5s}{\mathfrak{r}{\cal H}}+ 5s- b_e\right]
\times \vs
&\times& \left[\Psi +\int_0^{\mathfrak{r}}d\mathfrak{r} (\Phi'+\Psi')\right] + \frac{2-5s}{\mathfrak{r}}\int d\mathfrak{r} (\Phi+\Psi) \, ,
\ea
where a prime indicates a derivative w.r.t. conformal time.
Here ${\bf v}$ is the peculiar velocity, $\Phi$ and $\Psi$ are the gravitational potentials, $\delta$ is the density contrast in comoving gauge, ${\cal H} = aH$ is the conformal Hubble parameter;
$\mathfrak{r}$ is the conformal distance on the light cone, $\mathfrak{r}(z)=\tau_0-\tau(z)$, with $\tau(z)$ being the conformal time. \\

Errors in the angular correlations are computed as in Equation~\eqref{eq:err-clgt}, and they are proportional to the amplitude of the correlations, cosmic variance and the number density of objects. Hence, given a galaxy survey, the best chance to detect the effect of tensor perturbations would be, as pointed out in~\citealt{Jeong:2012}, by looking at the effects on radial cross-bin correlations.
In the context of the full {\it observed} galaxy correlations, radial correlations are dominated by cosmic magnification terms~\citep{Raccanelli:radial}, that can be orders of magnitude larger than the newtonian prediction~\citep{Raccanelli:2015GR}.
However, magnification terms depend on the magnification bias $s$; therefore, if one selects a sample of galaxies with $s=0.4$, it is clear from Equation~\eqref{eq:cosmag} that the magnification contribution is canceled.
Moreover, as mentioned above, tensor perturbation effects on LSS are larger closer to the time of emission, so that high-redshift bins are the most useful.
Therefore, radio continuum surveys with the CBR technique provides the ideal scenario.

\begin{figure}
\centering
\includegraphics[width=\columnwidth]{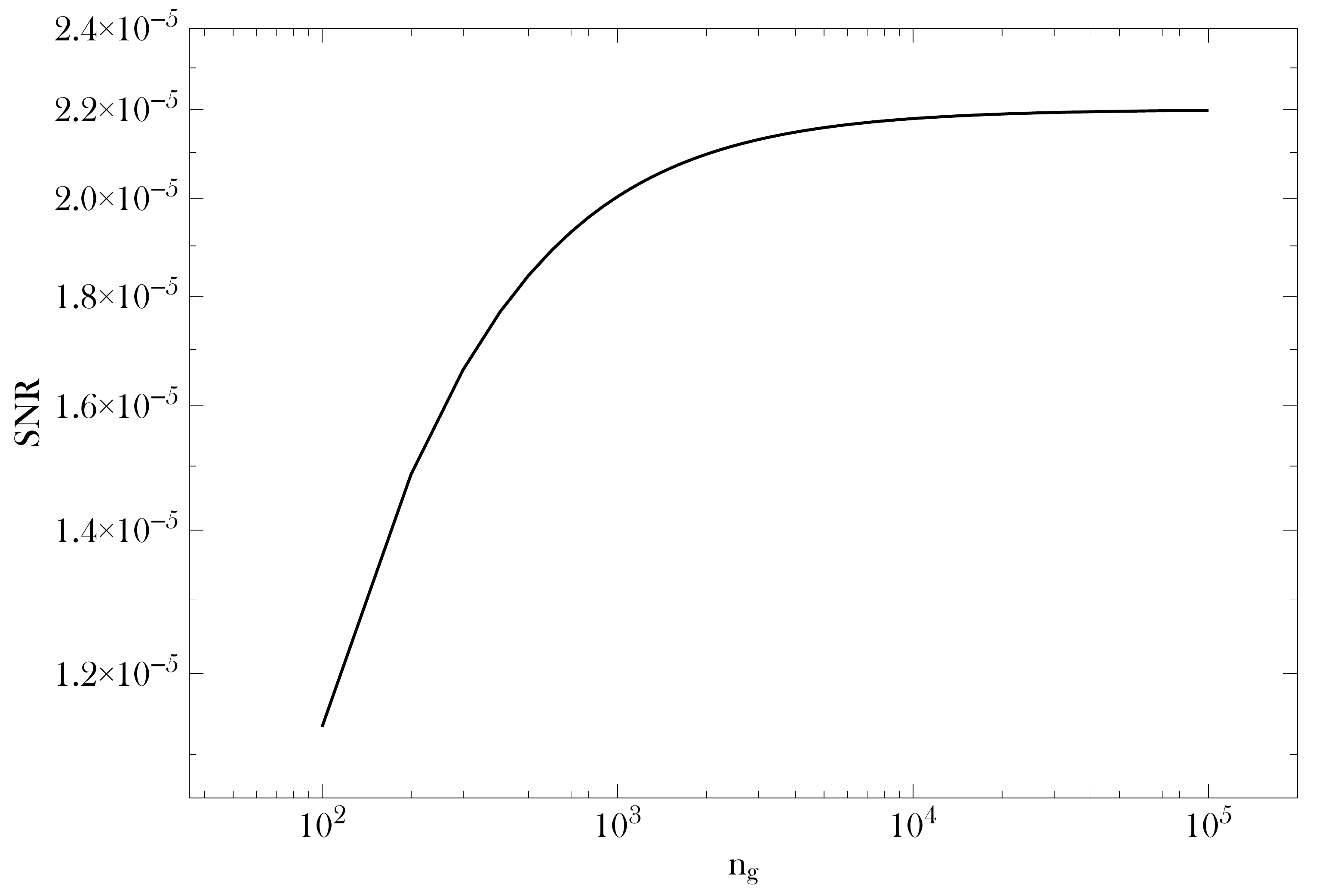}
\caption{Signal-to-noise ratio for tensor perturbation contributions (assuming r=0.09) to the {\it observed} galaxy 2-point correlation function.}
\label{fig:snr_clu}
\end{figure}

Unfortunately, even in the most optimistic cases with high-z sources and small shot noise, tensor perturbations remain undetectable. In Figure~\ref{fig:snr_clu} we show the signal-to-noise ratio, defined as
$(C_\ell^{\rm tot}-C_\ell^{\mathfrak{t}})/\sigma_{C_\ell}$, for the case of the $S_{\rm lim} = 1 \,n$Jy continuum survey.
As pointed out above, radial cross-bin correlations at high-z are the most promising ways to detect this signal, so we computed forecasts by using radio continuum surveys with 5 bins centered at $z=\{2, 3, 4, 5, 6\}$. The plot shows the SNR for the first 20 multipoles $\ell$, for different cross-bin radial correlations.

Our results show that even in the case of negligible shot noise and high-z correlations, the total SNR achievable will not reach $0.0001$.
While in principle the multi-tracer technique and optimization on the number of bins, galaxy populations, bias, and other parameters can significantly increase the SNR, it appears clear that it is very unlikely any survey built in the foreseeable future could allow a robust measurement using this observable.

\subsubsection{Clustering fossils}
In principle, another way of measuring effects from primordial GWs has been developed in~\cite{Jeong:2012CF, Dai:2016}, where it is shown how to search a galaxy survey for the imprint of primordial scalar, vector, and tensor fields. However, such effects require an extremely futuristic galaxy survey and measurements in the highly non-linear regime or higher order correlations; in this paper we do not focus on detailed modeling of non-linearities nor higher order correlations, hence we leave a careful investigation of this formalism to a future work.

\subsubsection{Weak lensing}
Now we turn our focus on effects of tensor perturbations on the correlation of galaxy ellipticities, referring again to the ``Cosmic Rulers'' formalism, and in particular we follow~\cite{Schmidt:2012, Schmidt:2014}.
In the unperturbed case, one would expect galaxy ellipticities to be uncorrelated, but unlike in the case of the galaxy correlation function, shear is a tensorial quantity, and thus there is a possible intrinsic contribution correlated with the GW background; this can be thought as analogous to the intrinsic alignment effect present for scalar perturbations.
Scalar perturbations contribute only to the E-mode component (at linear order), while tensor perturbations also contribute to the B-mode.
Hence, the curl-mode of the correlation of galaxy ellipticities can be used to detect a stochastic gravitational wave background.
In the rest of this Section we summarize what is the effect of a background of GWs to the correlation of ellipticities, and study what are the minimum experiment configurations required to detect the effect of such background, assuming $r=0.09$.

Once again, we consider angular correlations, and compute them with Equation~\eqref{eq:CABl}; in the case of lensing from GWs, we obtain
the angular power spectra of $E$- and $B$-modes of the shear induced by tensor modes by using:
\ba
\label{eq:Clshear1}
C_\gamma^{\mathfrak{X}\mathfrak{X}}(\ell) = \frac1{2\pi} 
\int k^2 dk\: P_{T0}(k) |F_l^{\gamma \mathfrak{X}}(k)|^2, \,\,\, {\bf \bar{\mathfrak{X}}} = \{E,\,B\} \, ,
\ea
where ${F_l^{\gamma \bf \bar{\mathfrak{X}}}}$ is the kernel that accounts for the effect of the gradient and curl components, ${\bf \bar{\mathfrak{X}}} = \{E,\,B\}$, that can be written as:
\begin{widetext}
\ba
F_l^{\gamma E}(k) &=&
- \frac{1}{4} 
\bigg\{ T_T(k,\eta_0) \left(
-\frac1{\mathfrak{q}_0^2} \left[ (2\mathfrak{q}_0^2 - \ell^2 - 3\ell-2) j_\ell(\mathfrak{q}_0) + 2\mathfrak{q}_0 j_{\ell+1}(\mathfrak{q})\right]
\right) + 
 \left(1 - \frac{2}{3} \frac{C_1\rho_{c_0}}{H_0^{2}\tilde a^{2}} \left\{\partial_{\tilde\eta}^2 
+ \tilde a \tilde H \partial_{\tilde{\eta}}\right\}\right)
T_T(k,\tilde\eta) - \vs
&-& \frac1{\tilde{\mathfrak{q}}^2} \left[ (2\tilde{\mathfrak{q}}^2 - \ell^2 - 3\ell -2) j_\ell(\tilde{\mathfrak{q}}) + 2\tilde{\mathfrak{q}} j_{\ell+1}(\tilde{\mathfrak{q}})\right]
\bigg\}
+ \int_0^{\chit} \frac{d\chi}{\chi}
\bigg[ 
-\frac14 (\ell+2)(\ell-1) \left[(\ell+1)(\ell-2) \frac{j_\ell(\mathfrak{q})}{\mathfrak{q}^2} + 2 \frac{j_{\ell+1}(\mathfrak{q})}{\mathfrak{q}}\right] + \vs
&+& \frac{\chi}{\chit} 
\frac14 \frac{(\ell+2)!}{(\ell-2)!} \frac{j_\ell(\mathfrak{q})}{\mathfrak{q}^2}
\bigg]
T_T(k, \eta_0-\chi) 
\times \left[2(\ell-1) \frac{j_\ell(\mathfrak{q})}{\mathfrak{q}} - 2j_{\ell-1}(\mathfrak{q}) \right] \, ; \vs
F_l^{\gamma B}(k) &=& 
- \frac{1}{4} 
\bigg\{ T_T(k,\eta_0) 
 \left[2(\ell-1) \frac{j_\ell(\mathfrak{q}_0)}{\mathfrak{q}_0} - 2j_{\ell-1}(\mathfrak{q}_0) \right] + \left(1 - \frac{2}{3} \frac{C_1\rho_{c_0}}{H_0^{2}\tilde a^{2}} \left\{\partial_{\tilde\eta}^2 + \tilde a \tilde H \partial_{\tilde\eta}\right\}\right) \times \vs
&\times& T_T(k,\tilde\eta)
2 \left[(\ell-1) \frac{j_\ell(\tilde{\mathfrak{q}})}{\tilde{\mathfrak{q}}} - j_{\ell-1}(\tilde{\mathfrak{q}}) \right]
\bigg\}
- \int_0^{\chit} \frac{d\chi}\chi
 \frac{(\ell-1)(\ell+2)}{2} \frac{j_\ell(\mathfrak{q})}{\mathfrak{q}} T_T(k, \eta_0-\chi) \, ,
\ea
\end{widetext}
where $\mathfrak{q}=k\chi$, and the constant of proportionality $C_1$ determines the magnitude of alignment and is usually of the order of $\sim 0.1$.

We compute the errors on the angular correlations by using~\citep{Schmidt:2012}:
\begin{equation}
\sigma_{C_{\ell}^{\mathfrak{X}\mathfrak{X}}} = \frac{\sigma_e^2}{2 \bar{n}_g \hksqrt{(2\ell+1)f_{\rm sky}}} \, ,
\end{equation}
where $f_{\rm sky}$ is the fraction of sky surveyed, $\sigma_e$ is the {\it rms} intrinsic ellipticity of galaxies, and $\bar{n}_g$
is the number of detected galaxies per steradian.

\begin{figure}
\centering
\includegraphics[width=\columnwidth]{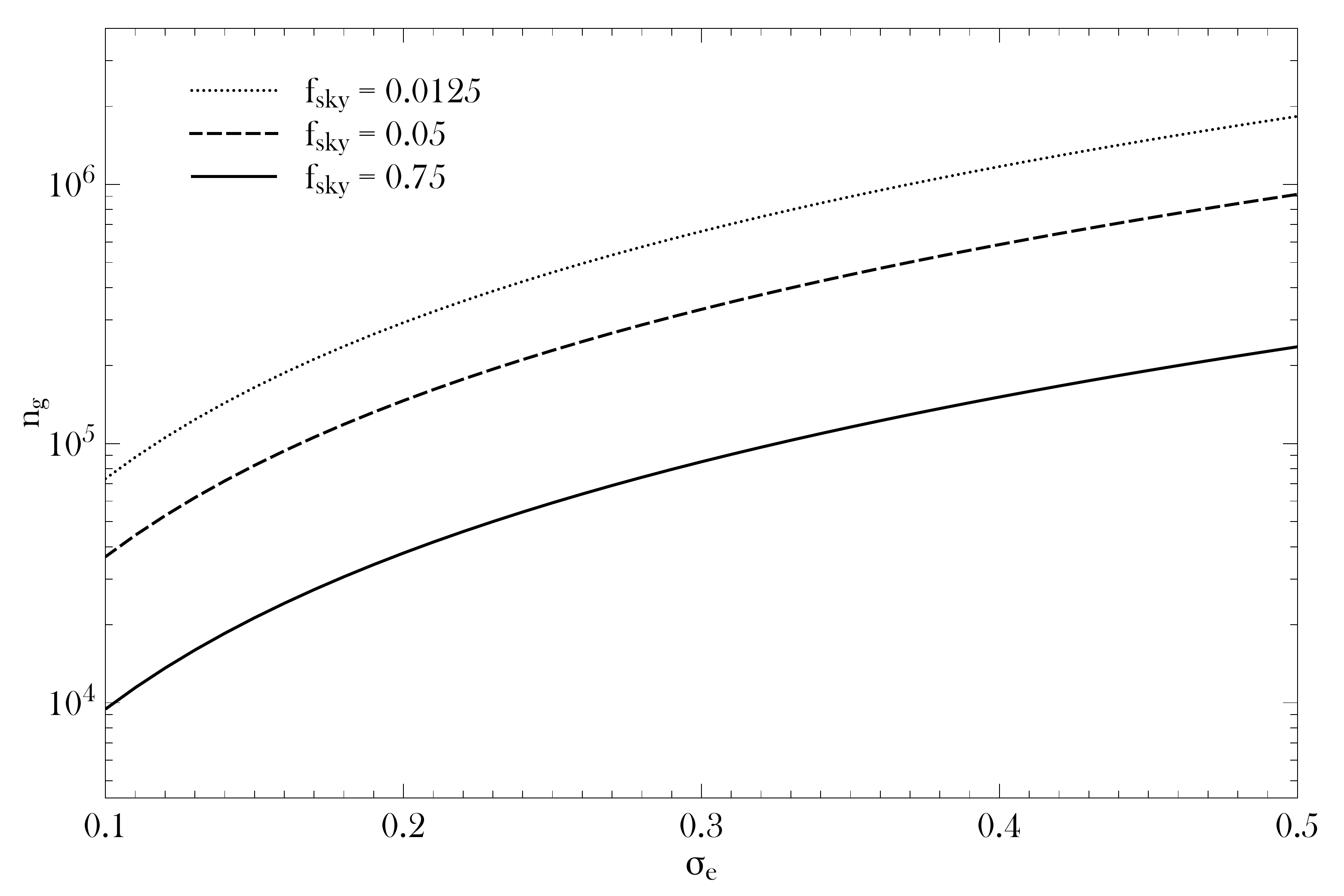}
\caption{Minimum number of galaxies per deg.$^2$ per bin needed to detect gravitational waves from inflation with $r=0.09$, using the correlation of galaxy ellipticities, as a function of intrinsic ellipticity error $\sigma_e$, for three different sky coverage values. See text for details.}
\label{fig:len1}
\end{figure}

The results are obtained by calculating the combined SNR of correlations of $C_\gamma^{BB}(\ell)$ at redshift bins centered at $z=\{2, 3, 4, 5, 6, 7, 8\}$, assuming an almost flat $N(z)$ as in the continuum 1 nJy case of Figure~\ref{fig:Nz_radio}. 
In Figure~\ref{fig:len1} we show the combination of $\sigma_e$ and $\bar{n}_g$ required to have a 1-$\sigma$ detection of the gravitational wave background with $r=0.09$.

We compute our results for three different survey areas, $f_{\rm sky}=\{0.0125, 0.05, 0.75\}$. Of course a smaller area surveyed would require a larger minimum number density, but very large number densities could be more easily achievable with dedicated deep surveys focusing on a smaller part of the sky.

In any case, for futuristic surveys as in Figure~\ref{fig:Nz_radio}, the most optimistic case used in this paper predicts around $10^4$ sources per sq. deg., meaning that in order to have such detection, a survey should observe a total of $\sim 10^5$ objects above $z=2$, with $\sigma_e \approx 0.1$ over 75\% of the sky, making it a very difficult target but somewhat more doable than the galaxy clustering case.

\section{Conclusions}
\label{sec:conclusions}
This work analyzed a few possible ways to achieve this goal: by cross-correlating radio galaxy catalogs with GW maps it is possible to determine properties of the progenitors of merging black hole binaries and forecast how the magnification of GWs by foreground radio sources allows to test models that explain cosmic acceleration.
Furthermore, it is in principle possible to detect the effects of a background of gravitational waves, by measuring angular motion and large scale correlations of galaxy distribution and lensing.

By using simulated catalogs resembling planned instruments, it was possible to show that the angular galaxy-GW cross-correlation can set stringent limits on properties of the progenitors of binary black holes, testing formation models and the possibility that primordial black holes are in fact the dark matter.

Radial cross-correlations can be used to detect the magnification bias of GWs that are lensed by low-redshift galaxies, and future laser interferometers, paired with forthcoming radio galaxies, can provide constraints on dynamical dark energy and modified gravity parameters that are competitive with the ones obtained with galaxy surveys alone and in combination with the CMB.

On the other hand, the detection of a stochastic gravitational wave background on galaxy position and distribution presents a much greater challenge. Tensor perturbation effects on galaxy clustering remain orders of magnitude below errors in measurements of galaxy power spectra, even in the case of very futuristic galaxy surveys.
The situation is slightly more optimistic for the correlation of lensing effects: in the case of a deep, full-sky survey with very precise shape measurements, it will be possible to measure the SGWB (provided that the tensor-to-scalar ratio $r$ is not negligibly small).

In any case, any measurements of gravitational wave effects coming from radio galaxy surveys or their correlation with GW detectors, would represent a valuable cross-check of other measurements and potentially provide new insights about cosmological models of current interest.


In summary, radio galaxy surveys can be used to provide information useful for gravitational wave astronomy and contribute studying the Universe in a new and complementary way.



\section*{Acknowledgments}
AR is indebted to Ely Kovetz for encouraging the development of this project and carefully reading the manuscript.
AR would like to thank Yacine Ali-Ha\"{i}moud, Ilias Cholis, Julian Munoz, Gianfranco de Zotti, Maria Chiara Guzzetti, Raul Jimenez and Licia Verde for commenting on the manuscript; Daniele Bertacca, Stefano Camera, Marc Kamionkowski, Sabino Matarrese, Prina Patel for useful discussions.
AR is supported by the John Templeton foundation.



\bibliography{GW_Radio_Sub}

\end{document}